\pdfoutput=1
\documentclass[a4paper, 11pt, oneside]{article}

\usepackage{epsf,amsmath,amssymb,graphicx,dcolumn}
\usepackage{scalefnt,cite,hyperref,bbold}
\usepackage{cleveref,etoolbox,color,soul}
\usepackage{listings,booktabs}
\usepackage{epstopdf}
\usepackage{amsfonts}

\newcommand{\half}{\tfrac{1}{2}}
\newcommand{\nc}{\newcommand}
\newcommand{\mchsq}{M_{H^\pm}^2}
\newlength{\absize}
\newcommand{\pcal}{{\cal P}}

\renewcommand{\Re}{\mbox{Re\thinspace}}
\renewcommand{\Im}{\mbox{Im\thinspace}}

\nc{\beq}{\begin{equation}}  \nc{\eeq}{\end{equation}}
\nc{\bea}{\begin{eqnarray}}  \nc{\eea}{\end{eqnarray}}

\def\thetaW{{\theta}_{\rm W}}
\def\lcal{{\cal L}}

\def\Tr{{\rm Tr\,}}

\numberwithin{equation}{section}

\newcounter{notecount}
\begin{document}
\begin{flushleft}
\today
\end{flushleft}

\long\def\symbolfootnote[#1]#2{\begingroup%
\def\thefootnote{\fnsymbol{footnote}}\footnote[#1]{#2}\endgroup}

\vspace{0.1cm}

\begin{center}
\Large\bf\boldmath
New Symmetries of the Two-Higgs-Doublet Model
\unboldmath
\end{center}
\vspace{0.05cm}
\begin{center}
P. M. Ferreira$^{a,b}$,
B. Grzadkowski$^c$,
O. M. Ogreid$^d$,
P. Osland$^e$\symbolfootnote[0]{\begin{flushleft}Electronic addresses:
pmmferreira@fc.ul.pt, bohdan.grzadkowski@fuw.edu.pl, omo@hvl.no, Per.Osland@uib.no.
\end{flushleft}}\\[0.4cm]
{\small
{\sl${}^a$Instituto  Superior  de  Engenharia  de  Lisboa,  Portugal}\\[0.2em]
{\sl${}^b$Centro  de  F{\'i}sica  Te{\'o}rica  e  Computacional,  Universidade  de  Lisboa,  Portugal}\\[0.2em]
{\sl${}^c$Faculty of Physics, University of Warsaw, Pasteura 5, 02-093 Warsaw, Poland}\\[0.2em]
{\sl${}^d$Western Norway University of Applied Sciences, Postboks 7030, N-5020 Bergen, Norway}\\[0.2em]
{\sl${}^e$Department of Physics and Technology,
University of Bergen, Postboks 7803, N-5020 Bergen, Norway}
}
\end{center}

\begin{abstract}
\noindent
The Two Higgs Doublet Model invariant under the gauge group $SU(2)\times U(1)$ is known to have six
additional global discrete or continuous symmetries of its scalar sector. We have discovered regions of parameter space of the model which
are basis and renormalization group invariant to all orders of perturbation theory in the scalar and gauge sectors, but correspond
to none of the hitherto considered symmetries.
We therefore identify seven new symmetries of the model and discuss their phenomenology. Soft symmetry breaking is required
for some of these models so that electroweak symmetry breaking can occur. We show that, at least at the two-loop level, it
is possible to extend some of these symmetries to include fermions.
\end{abstract}

\setcounter{footnote}{0}

\section{Introduction}

The Two-Higgs-Doublet Model (2HDM) is one of the more popular extensions of the Standard Model (SM) of particle physics.
It was introduced by Lee in 1973~\cite{Lee:1973iz} to provide an additional source of CP violation, thus attempting
to explain the overwhelming prevalence of matter over antimatter in the universe. In its simplest form, the model
has the same gauge symmetries as the SM, same fermionic content -- but instead of a single $SU(2)$ spin-0 doublet,
the 2HDM has two, $\Phi_1$ and $\Phi_2$. The model has a rich phenomenology with a scalar spectrum comprising three neutral and one charged
elementary spin-0 states. Different versions of the 2HDM allow for the possibility of spontaneous CP-violation; provide
dark matter candidates whose stability is guaranteed by a discrete symmetry; may have tree-level flavour changing neutral
currents (FCNCs) mediated by neutral scalars; may have sizeable contributions to flavour physics. For a review,
see for instance~\cite{Branco:2011iw}.

The scalar potential of the SM is characterized by 2 real, independent parameters, out of which one obtains
the value of the Higgs field vacuum expectation value (vev), $v = 246$ GeV, and the Higgs mass, $m_h \simeq 125$ GeV.
For the 2HDM, however, the scalar potential is much more complex: the most general 2HDM has a potential with
11 independent real parameters~\cite{Davidson:2005cw}. Simultaneously, that model has scalar-mediated FCNC,
which experimentally are known to be very constrained -- this arises because, in the most general 2HDM, both doublets
couple to fermions of the same  electric charge. For that reason, in 1976 a discrete $Z_2$ symmetry
was proposed to eliminate those FCNCs, so that fermions of the same charge (charged leptons, up-like and
down-like quarks) are made to couple to a single Higgs doublet~\cite{Glashow:1976nt, Paschos:1976ay}.
Along the way the number of free scalar parameters is reduced to 7, and thus the predictivity of the model is increased.
This $Z_2$ symmetry required invariance of the lagrangian under a transformation for which one of the doublets
changes sign while the other remains unchanged, for instance $\Phi_2 \rightarrow -\Phi_2$. In another example,
Peccei and Quinn~\cite{Peccei:1977hh} observed that a 2HDM endowed with a continuous global $U(1)$ symmetry
was a possible solution to the strong CP problem -- and in that model the number of free parameters of the scalar potential
is 6. The Peccei-Quinn symmetry may be obtained by requiring invariance under a transformation like $\Phi_2 \rightarrow e^{i\alpha}
\Phi_2$, for an arbitrary real phase $\alpha$. These are examples of unitary symmetry transformations between the doublets,
sometimes called {\em Higgs family symmetries}. Anti-unitary ones, which transform doublets into a linear combination of their
complex conjugates, or more precisely their CP conjugates, are also possible, and are called {\em generalized CP symmetries}.
These two types of field transformations -- unitary and anti-unitary -- leave invariant the doublets' kinetic terms, and it
has been shown~\cite{Ivanov:2005hg,Ivanov:2006yq} that, for the $SU(2)\times U(1)$ invariant scalar potential, there only
{\em six} possible symmetries. Since in the 2HDM both doublets have the same quantum numbers, any linear combination
thereof which preserves the kinetic terms is equally acceptable. This freedom to choose a basis of scalar fields may
mask the form of the symmetries, so that it may seem there are more than six of them. In fact a basis-independent analysis
shows that indeed, only six different symmetries -- and therefore six different versions of the 2HDM, with different numbers
of free parameters and possible phenomenology -- are allowed, when one considers all possible doublet transformations
which preserve the kinetic terms and gauge symmetries.

A fingerprint of continuous symmetries, from Noether's theorem~\cite{Noether:1918zz}, is the existence
of some quantities (charges) which are conserved  during the evolution of the system under its equations
of motion. Indeed, for each of the six symmetries mentioned above (and explained in
greater detail in section~\ref{sec:sym}) certain relations between parameters of the 2HDM scalar potential are found to be
preserved under renormalization. Symmetry-constrained relations between the dimensionless couplings of the model will even
remain invariant to all orders of perturbation after spontaneous symmetry breaking of that symmetry has
occurred~\footnote{Finite contributions to those couplings from radiative corrections may spoil those relations, however.}.

In this paper we will investigate a curious situation in which we have been able to identify a region of 2HDM
parameter space characterized by specific relations between couplings which are not only basis invariant but also
left invariant under the renormalization group (RG) -- and which do not correspond to any of the six aforementioned symmetries.
In terms of the most usual notation used to write the 2HDM scalar potential, these conditions are
\beq
m^2_{11} + m^2_{22} = 0 \;\,,\,\; \lambda_1=\lambda_2 \;\,,\,\;  \lambda_7=-\lambda_6  \,.
\eeq
Using arguments of basis invariance,
we will show how this specific region of the 2HDM parameter space remains invariant under renormalization to
all orders of perturbation theory, not considering fermions. We were unable to extend the all-order argument to the Yukawa
sector, but will show, via an explicit calculation, that the relations between parameters we have found remain invariant
at least to two loops when fermions are taken into account. We therefore conclude that, at least to two-loop order,
the specific relations between couplings which we found are invariant under renormalization when the whole lagrangian
is taken into account. Indeed, it could be that invariance at one-loop would be the consequence of some unphysical fine-tuning,
but to see that those relations between couplings remain valid even when two-loop contributions are taken into account
suggests that invariance to all orders is a strong possibility. To put things into perspective, consider that multi Higgs
doublets are many times studied under the so-called ``custodial symmetry", which is an approximate symmetry of the lagrangian.
The scalar potential can be made
invariant implying a specific mass spectrum for scalars. Invariance of the kinetic terms
under custodial symmetry would imply equal masses for the W and Z bosons, and is
therefore broken by the presence of the gauge coupling constant corresponding to the
hypercharge gauge group.
 It is also broken by Yukawa interactions, namely by the fact that up-type and down-type
quarks have different masses. Therefore, custodial symmetry relations are not preserved under radiative corrections even at
the one-loop level.

However, we cannot find what specific field transformation yields these RG-invariant conditions. We know that it cannot be
a unitary or anti-unitary transformation on the doublets. We have identified a transformation on scalar bilinears --
quadratic combinations of scalar doublets which are gauge invariant -- which seemingly produces exactly the region
of parameter space we are interested in, but not only such a transformation is impossible to reproduce on the basis of
operations upon doublets, it does not seem possible to extend it to the gauge sector, let alone the Yukawa one. Nonetheless,
though ignorant of the transformation on fields which produces this RG invariant region, we will nevertheless refer to it as
being produced by a {\em symmetry}, which we call the {\em $r_0$ symmetry}. It is possible
to combine the $r_0$ symmetry with the other six known 2HDM symmetries and find new models, which boast (new) combinations of parameters
which are RG-invariant to all orders, and quite interesting phenomenologies, including, for specific models: existence of explicit CP violation;
impossibility of spontaneous breaking of a $Z_2$ symmetry or CP violation; mass degeneracy of neutral scalars; and
no decoupling limit possible when the $r_0$ symmetry holds. While extending the $r_0$ symmetry to the fermion sector we will
prove that the Yukawa matrices found obeying previously known 2HDM symmetries (to wit, the symmetries called CP2 and CP3) also
preserve the new conditions among parameters characteristic of the new symmetry to at least two-loop order.

This paper is structured as follows: in section~\ref{sec:2hdm} we review the 2HDM, with an emphasis on basis transformations,
the bilinear formalism, the known symmetries of the model and the model's one-loop
renormalization group equations, which will be the stepping stone for our reasoning.
In section~\ref{sec:news} we will show how the set of relations between 2HDM parameters shown above is preserved under renormalization
at the one-loop level. We will then demonstrate how, considering only the scalar and gauge sectors of the model to begin with,
that invariance is indeed an all-order result, using arguments of basis invariance and dimensional analysis to perform an analysis
of the model's $\beta$-functions at an arbitrary number of loops. We then provide a heuristic interpretation of this symmetry
using the bilinear formalism, which shows how the desired conditions upon the parameters may be obtained via a sign change in
one of the bilinears in a formal manner, which justifies the name  $r_0$ symmetry we chose. We then combine the $r_0$ symmetry
conditions with those of the known 6 2HDM symmetries and list 7 new possible symmetries of the model. Some of those lead to
vanishing quadratic terms and must be softly broken. In section~\ref{sec:sphen} we analyse the phenomenology of the scalar
sector of each of the symmetries considered, including soft breaking terms when necessary or interesting. Section~\ref{sec:ferm}
sees us tackling the fermion sector and arguing that the CP2 or CP3 Yukawa textures would adequately preserve the $r_0$ symmetry
relations between quartic couplings to all orders, and showing by means of an explicit $\beta$-function calculation, that
those same Yukawa structures also preserve, at least to two-loop order, the relation $m^2_{11} + m^2_{22} = 0$. An overview
of our results and conclusions are drawn in section~\ref{sec:conclusions}.

\section{The Two-Higgs Doublet Model}
\label{sec:2hdm}

The 2HDM is one of the simplest extensions of the SM, wherein one considers two $SU(2)$ doublets with hypercharge one
 instead of just one doublet. In the following we will briefly review the basic aspects of a useful formalism
to understand the structure of the scalar sector of the model, and the global symmetries one can impose
on it.

\subsection{The scalar potential}

The most general scalar potential involving two hypercharge $Y = 1$ scalar doublets invariant
under the electroweak gauge group $SU(2)_L\times U(1)_Y$ is given by
\bea
V &=& m_{11}^2\Phi_1^\dagger\Phi_1+m_{22}^2\Phi_2^\dagger\Phi_2
-[m_{12}^2\Phi_1^\dagger\Phi_2+{\rm h.c.}]+\half\lambda_1(\Phi_1^\dagger\Phi_1)^2
+\half\lambda_2(\Phi_2^\dagger\Phi_2)^2
+\lambda_3(\Phi_1^\dagger\Phi_1)(\Phi_2^\dagger\Phi_2)\nonumber\\[8pt]
&&\quad
+\lambda_4(\Phi_1^\dagger\Phi_2)(\Phi_2^\dagger\Phi_1)
+\left\{\half\lambda_5(\Phi_1^\dagger\Phi_2)^2
+\big[\lambda_6(\Phi_1^\dagger\Phi_1)
+\lambda_7(\Phi_2^\dagger\Phi_2)\big]
\Phi_1^\dagger\Phi_2+{\rm h.c.}\right\}\,,
 \label{eq:pot}
\eea
where, other than $m^2_{12}$ and $\lambda_{5,6,7}$, all parameters are real. An alternative notation uses
four gauge-invariant bilinears constructed from the
doublets\cite{Velhinho:1994np,Nagel:2004sw,Ivanov:2005hg,Ivanov:2006yq,Ivanov:2007de,Maniatis:2006fs,Maniatis:2006jd,
Nishi:2006tg,Nishi:2007nh,Nishi:2007dv,Maniatis:2007vn},
\beq
\begin{array}{rcl}
r_0 &=&
\frac{1}{2}
\left( \Phi_1^\dagger \Phi_1 + \Phi_2^\dagger \Phi_2 \right),
\\*[2mm]
r_1 &=&
\frac{1}{2}
\left( \Phi_1^\dagger \Phi_2 + \Phi_2^\dagger \Phi_1 \right)
= \mbox{Re}\left( \Phi_1^\dagger \Phi_2 \right),
\\*[2mm]
r_2 &=&
- \frac{i}{2}
\left( \Phi_1^\dagger \Phi_2 - \Phi_2^\dagger \Phi_1 \right)
= \mbox{Im} \left( \Phi_1^\dagger \Phi_2 \right),
\\*[2mm]
r_3 &=&
\frac{1}{2}
\left( \Phi_1^\dagger \Phi_1 - \Phi_2^\dagger \Phi_2 \right).
\end{array}
\label{eq:rs}
\eeq
In terms of these quantities, then, the potential of eq.~\eqref{eq:pot} may be written as
\beq
V \,=\,M_\mu\,r^\mu\,+\,\Lambda_{\mu\nu}\,r^\mu\,r^\nu\,,
\eeq
where we use a Minkowski-like formalism to define the 4-vectors
\bea
r^\mu &=& (r_0\,,\,r_1\,,\,r_2\,,\,r_3) \,=\, (r_0\,,\,\vec{r})\,, \nonumber \\
M^\mu &=& \left(m^2_{11} + m^2_{22}\,,\, 2\mbox{Re}(m^2_{12})
\,,\, -2\mbox{Im}(m^2_{12})\,,\,m^2_{22} - m^2_{11}\right)\,=\, (M_0\,,\,\vec{M})\,,
\label{eq:defbil}
\eea
as well as the tensor
\begin{align}
	\Lambda^{\mu\nu} & = \begin{pmatrix} \Lambda_{00} & \vec{\Lambda} \\
\vec{\Lambda}^T & \Lambda \end{pmatrix}
	=
	\begin{pmatrix}
		\frac{1}{2}(\lambda_1 + \lambda_2) + \lambda_3 &
	 -\mbox{Re}\left(\lambda_6 + \lambda_7\right) &
	\mbox{Im}\left(\lambda_6 + \lambda_7\right) &
		\frac{1}{2}(\lambda_2 - \lambda_1) \\
		-\mbox{Re}\left(\lambda_6 + \lambda_7\right) &
		\lambda_4 + \mbox{Re} \left( \lambda_5\right) &
		- \mbox{Im} \left( \lambda_5 \right) &
		\mbox{Re}\left(\lambda_6 - \lambda_7\right)
	\\
	\mbox{Im} \left( \lambda_6 + \lambda_7\right)&
	- \mbox{Im} \left( \lambda_5 \right) &
	\lambda_4 - \mbox{Re} \left( \lambda_5\right) &
	- \mbox{Im} \left( \lambda_6 - \lambda_7\right) \\
	\frac{1}{2}(\lambda_2 - \lambda_1) &
	\mbox{Re}\left( \lambda_6 - \lambda_7\right) &
	- \mbox{Im} \left( \lambda_6 - \lambda_7 \right) &
	\frac{1}{2}(\lambda_1 + \lambda_2) - \lambda_3
	\end{pmatrix}\,.
\label{eq:Lambda}
\end{align}
For future convenience, we defined the singlet $\Lambda_{00}$ and the vector $\vec{\Lambda}$ as
\beq
\Lambda_{00}\,=\,\frac{1}{2}(\lambda_1 + \lambda_2) + \lambda_3\;\;\; , \;\;\;
\vec{\Lambda}\,=\,\left(-\mbox{Re}\left(\lambda_6 + \lambda_7\right) \,,\,
	\mbox{Im}\left(\lambda_6 + \lambda_7\right) \,,\,
		\frac{1}{2}(\lambda_2 - \lambda_1)\right)\,,
\label{eq:L0Lv}
\eeq
and therefore the matrix $\Lambda$ from eq.~\eqref{eq:Lambda} is the right-bottom $3\times 3$ block
within the $\Lambda^{\mu\nu}$ tensor above.

\subsection{Basis transformations}
\label{sec:basis}

Since both doublets have exactly the same quantum numbers, there is nothing {\em a priori} that
distinguishes one from the other -- thus any linear combination of the two that preserves the kinetic
terms of the theory should yield the same physics. Specifically, if one considers a new set of doublets
$\{\Phi_1^\prime\,,\,\Phi_2^\prime\}$ related to the first by $\Phi_a^\prime\,=\,U_{ab}\,\Phi_b$, for any
unitary $2\times 2$ matrix $U$, the model, and all physics thereof originating, is left invariant.
These are called {\em basis transformations}, and the parameters of the potential will in general change
from basis to basis. If we parameterize the matrix $U$ as
\beq
U = \left( \begin{array}{cc}
e^{i \chi} c_\psi & e^{i \left( \chi - \xi \right)} s_\psi \\
- e^{i \left( \xi - \chi \right)} s_\psi & e^{- i \chi} c_\psi
\end{array} \right),
\eeq
where we have defined $c_x = \cos x$ and $s_x = \sin x$,
we obtain relations between the parameters of the potential in the new basis as a function of
those in the original one and the angles and phases which form $U$~\cite{Gunion:2005ja,Branco:2011iw}:
\begin{subequations}
\bea
{m_{11}^2}^\prime &=&
m_{11}^2 c_\psi^2 + m_{22}^2 s_\psi^2
- \mbox{Re}\left( m_{12}^2 e^{i \xi} \right) s_{2 \psi},
\label{eq:M11}
\\
{m_{22}^2}^\prime &=&
m_{11}^2 s_\psi^2 + m_{22}^2 c_\psi^2
+ \mbox{Re}\left( m_{12}^2 e^{i \xi} \right) s_{2 \psi},
\label{eq:M22}
\\
{m_{12}^2}^\prime &=&
e^{i \left( 2 \chi - \xi \right)} \left[
\frac{1}{2} \left( m_{11}^2 - m_{22}^2 \right) s_{2 \psi}
+ \mbox{Re}\left( m_{12}^2 e^{i \xi} \right) c_{2 \psi}
+ i \mbox{Im} \left( m_{12}^2 e^{i \xi} \right)
\right],
\label{eq:M12}
\\
\lambda_1^\prime &=&
\lambda_1 c_\psi^4 + \lambda_2 s_\psi^4
+ \frac{1}{2} \lambda_{345} s_{2 \psi}^2
+ 2 s_{2 \psi} \left[
c_\psi^2 \mbox{Re}\left( \lambda_6 e^{i \xi} \right)
+ s_\psi^2 \mbox{Re}\left( \lambda_7 e^{i \xi} \right)
\right],
\label{eq:L1}
\\
\lambda_2^\prime &=&
\lambda_1 s_\psi^4 + \lambda_2 c_\psi^4
+ \frac{1}{2} \lambda_{345} s_{2 \psi}^2
- 2 s_{2 \psi} \left[
s_\psi^2 \mbox{Re}\left( \lambda_6 e^{i \xi} \right)
+ c_\psi^2 \mbox{Re}\left( \lambda_7 e^{i \xi} \right)
\right],
\label{eq:L2}
\\
\lambda_3^\prime &=& \lambda_3
+ \tfrac{1}{4} s_{2 \psi}^2 \left(
\lambda_1 + \lambda_2 - 2 \lambda_{345} \right)
- s_{2 \psi} c_{2 \psi}
\mbox{Re}\left[ \left( \lambda_6 - \lambda_7 \right) e^{i\xi} \right],
\label{eq:L3}
\\
\lambda_4^\prime &=&
\lambda_4
+ \frac{1}{4} s_{2 \psi}^2
\left( \lambda_1 + \lambda_2 - 2 \lambda_{345} \right)
- s_{2 \psi} c_{2 \psi}
\mbox{Re}\left[ \left( \lambda_6 - \lambda_7 \right) e^{i \xi} \right],
\label{eq:L4}
\\
\lambda_5^\prime &=&
e^{2 i \left( 2 \chi - \xi \right)} \left\{
\frac{1}{4} s_{2 \psi}^2
\left( \lambda_1 + \lambda_2 - 2 \lambda_{345} \right)
+ \mbox{Re}\left( \lambda_5 e^{2 i \xi} \right)
+ i c_{2 \psi} \mbox{Im} \left( \lambda_5 e^{ 2 i \xi}\right)
\right. \nonumber \\*[1mm] & & \left.
- s_{2 \psi} c_{2 \psi}
\mbox{Re}\left[ \left( \lambda_6 - \lambda_7 \right) e^{i \xi} \right]
-i s_{2 \psi} \mbox{Im}
\left[ \left( \lambda_6 - \lambda_7 \right) e^{i\xi} \right]
\right\},
\label{eq:L5}
\\
\lambda_6^\prime &=&
e^{i \left( 2\chi - \xi \right)} \left\{
- \frac{1}{2} s_{2 \psi} \left[
\lambda_1 c_\psi^2 - \lambda_2 s_\psi^2 - \lambda_{345} c_{2 \psi}
- i \mbox{Im} \left( \lambda_5 e^{2 i \xi} \right)
\right]
\right. \nonumber \\*[1mm] & & \left.
+ c_\psi c_{3 \psi} \mbox{Re}\left( \lambda_6 e^{i \xi} \right)
+ s_\psi s_{3 \psi} \mbox{Re}\left( \lambda_7 e^{i \xi}\right)
+ i c_\psi^2 \mbox{Im} \left( \lambda_6 e^{i \xi} \right)
+ i s_\psi^2 \mbox{Im} \left( \lambda_7 e^{i \xi} \right)
\right\},
\label{eq:L6}
\\
\lambda_7^\prime &=&
e^{i \left( 2\chi - \xi \right)} \left\{
- \frac{1}{2} s_{2 \psi} \left[
\lambda_1 s_\psi^2 - \lambda_2 c_\psi^2 + \lambda_{345} c_{2 \psi}
+ i \mbox{Im} \left( \lambda_5 e^{2 i \xi} \right) \right]
\right. \nonumber \\*[1mm] & & \left.
+ s_\psi s_{3 \psi} \mbox{Re}\left( \lambda_6 e^{i \xi} \right)
+ c_\psi c_{3 \psi} \mbox{Re}\left( \lambda_7 e^{i \xi} \right)
+ i s_\psi^2 \mbox{Im} \left( \lambda_6 e^{i \xi} \right)
+ i c_\psi^2 \mbox{Im} \left( \lambda_7 e^{i \xi} \right)
\right\},
\label{eq:L7}
\eea
\end{subequations}
where for convenience we write
\beq
\lambda_{345} = \lambda_3 + \lambda_ 4
+ \mbox{Re}\left( \lambda_5 e^{2 i \xi} \right).
\eeq

Basis transformations are exceedingly simple to write in the bilinear formalism.
Defining the $3\times 3$ matrix $O(3)$ rotation matrix $R_{ij}(U) = \Tr\left(U^\dagger
\sigma_i U \sigma_j\right)/2$, where $\sigma_i$ ($i = 1,2,3$) are the Pauli matrices, we find that $\vec{r}$,
$\vec{M}$ and $\vec{\Lambda}$ transform as vectors for these basis changes, {\em i.e}
\bea
{\vec{r}\,}^\prime &=& R\,\vec{r}\,,\nonumber \\
\vec{M}^\prime &=& R\,\vec{M}\,,\nonumber \\
\vec{\Lambda}^\prime &=& R\,\vec{\Lambda}\,,
\eea
whereas $r_0$, $M_0$ and $\Lambda_{00}$ do not change under basis transformations -- they are {\em basis
invariants} -- and $\Lambda$ transforms as a $3\times 3$ matrix would under rotations, $\Lambda^\prime
= R \,\Lambda\,R^T$.

The most general potential of eq.~\eqref{eq:pot} has seemingly 14 independent real parameters, but in fact,
once basis freedom is taken into account (which allows one to choose a basis so as to eliminate several
parameters), the real number of independent real parameters is 11~\cite{Davidson:2005cw}. This may be seen in several ways,
but perhaps the simplest of those is using the bilinear formalism described above: since basis transformations
correspond, in this formalism, to $O(3)$ rotations, the matrix $R$ is characterized by 3 independent angles, which
can be used to ``rotate away" three of the 14 parameters of the potential. For instance, one can chose $R$ so as
to diagonalize the symmetric $3 \times 3$ $\Lambda$ matrix, thus eliminating three out of its six parameters.

It is also possible to express the kinetic terms in terms of bilinears, though the limitations of this formalism start to
appear. As explained in~\cite{Ivanov:2007de}, the scalar kinetic terms $T$ (excluding gauge interactions) may be written as
\beq
T \,=\,K^{\mu}\,\left(\partial_\alpha \Phi_i\right)^\dagger\,(\sigma_\mu)_{ij}\,\left(\partial^\alpha \Phi_j\right)\, ,
\label{eq:kin}
\eeq
where a sum on $i,j = 1,2$ is assumed, and the 4-vectors in this expression are $\sigma^\mu\,=\,
(\mathbb{1}\,,\,\sigma_i)$, with $\sigma_i$ the Pauli matrices, and $K^\mu \,=\,
(1\,,\,0\,,\,0\,,\,0)$. Though we write the bilinears and the potential in a Minkowski-like
formalism, we should not consider boost transformations of the 4-vectors or tensors considered:
in fact, such transformations would change $K^\mu$ in such a way that eq.~\eqref{eq:kin} would
no longer yield the correct kinetic terms for the scalar doublets.

\subsection{Global symmetries of the 2HDM}
\label{sec:sym}

One can impose global symmetries on the 2HDM potential of eq.~\eqref{eq:pot} to obtain models with different
and interesting phenomenology. Following the usual procedure, one takes scalar field transformations which preserve
their kinetic
terms, and there are two possibilities for that to occur: one may consider {\it Higgs-family symmetries}, where unitary
transformations mix both doublets,
\beq
\Phi_i\,\rightarrow\, \Phi^\prime_i\,=\, \sum_{j=1}^2\,U_{ij}\,\Phi_j\,
\label{eq:hf}
\eeq
where $U$ is a generic $2\times 2$ unitary matrix; the other possibility is anti-unitary field transformations,
\beq
\Phi_i\,\rightarrow\, \Phi^\prime_i\,=\, \sum_{j=1}^2\,X_{ij}\,\Phi^*_j\,
\label{eq:gcp}
\eeq
where once again $X\in U(2)$ is a generic matrix but now the transformed fields are combinations of
the complex conjugates of the original doublets. These are called {\it generalized CP (GCP) symmetries}.
The simplest example of a transformation like those of eq.~\eqref{eq:hf} is a simple $Z_2$ symmetry, with
one of the doublets changing sign, while the other remains the same,
\beq
\Phi_1\,\rightarrow \,\Phi_1\;\;\; , \;\;\;
\Phi_2\,\rightarrow \,-\,\Phi_2\,.
\label{eq:z2}
\eeq
This symmetry, when extended to the Yukawa sector, prevents the occurrence of tree-level flavour-changing
neutral currents (FCNC)~\cite{Glashow:1976nt,Paschos:1976ay}, and eliminates the $m^2_{12}$, $\lambda_6$ and
$\lambda_7$ couplings. And the simplest example of a symmetry like those of eq.~\eqref{eq:gcp} is
the ``standard" CP transformation, {\it i.e.} requiring invariance of the potential under the field
transformation
\beq
\Phi_i\,\rightarrow \,\Phi_i^*\,.
\label{eq:cp1}
\eeq
This symmetry, sometimes called CP1, yields a potential for which there exists a basis such that
all parameters are real, and the possibility of
CP-conserving vacua exists, as well as vacua with spontaneous CP violation -- unlike the most general potential
of eq.~\eqref{eq:pot}, where CP breaking is explicit.

In the bilinear formalism, both Higgs-family and GCP field transformations are represented as rotations
in the 3-dimensional space defined by the vector $\vec{r}$, namely
\beq \label{Eq:S-matrix}
\vec{r}\,\rightarrow \, \vec{r}^\prime\,=\,S\,\vec{r}\,,
\eeq
where $S\,\in\,O(3)$ defines a rotation of $\vec{r}$. When such rotations are proper ({\it i.e.,} $\det(S) = 1$)
we have a Higgs-family symmetry. Improper rotations ($\det(S) = -1$) yield GCP symmetries. Both types of
symmetries/rotations leave the value of $r_0$ invariant, because they arise from unitary or anti-unitary field
transformations\footnote{Indeed, there is a well-defined procedure to obtain the  matrix $S$ from the $U$
and $X$ matrices
defined in eqs.~\eqref{eq:hf} and~\eqref{eq:gcp},
see~\cite{Ivanov:2005hg,Ivanov:2006yq,Maniatis:2007vn,Nishi:2006tg,Branco:2011iw} for details.}. The two examples of symmetries
described above correspond to $S$ matrices given by
\beq
S_{Z_2}\,=\,\begin{pmatrix} -1 & 0 & 0\\
 0 & -1 & 0 \\
 0 & 0 & 1\end{pmatrix}\;\;\; , \;\;\;
S_{CP1}\,=\,\begin{pmatrix} 1 & 0 & 0\\
 0 & -1 & 0 \\
 0 & 0 & 1\end{pmatrix}\,.
\eeq
Given the freedom to change basis that the 2HDM scalar potential possesses, the same symmetry may look differently
in different bases, but its physical implications remain the same. For instance, on a different basis,
the $Z_2$ symmetry actually looks like a permutation symmetry $S_2$, where the field transformation corresponds
to an exchange between the doublet fields, $\Phi_1\leftrightarrow \Phi_2$. The resulting potential
looks different from the one mentioned above (now we would have $m^2_{11} = m^2_{22}$, $\lambda_1 =
\lambda_2$ and $\lambda_6 = \lambda_7$), but it is simply a basis change from the basis wherein
the $Z_2$ field transformation is given by eq.~\eqref{eq:z2}. Indeed, the matrix~(\ref{Eq:S-matrix}) for the $S_2$ transformation
is simply given by $S_{S_2} \,=\,\mbox{diag}(1\,,\,-1\,,\,-1)$, which is clearly obtained from $S_{Z_2}$ by a
permutation of axis. Such permutations correspond, in the bilinear formalism, to basis changes. In fact,
it may be shown \cite{Ivanov:2005hg,Ivanov:2006yq} that the $Z_2$ symmetry corresponds,  in an arbitrary basis,
to a parity transformation ({\it i.e.} a sign flip) on two of the three axis of the $\vec{r}$ vector. Likewise,
the CP1 symmetry will always be given by a parity transformation on a single of the three axis of this space, and there is no
physical distinction between a parity transformation on the first, second or third axis (these would correspond to transformations
on $\vec{r}$ such that $r_1 \rightarrow -r_1$, or $r_2 \rightarrow -r_2$ or $r_3 \rightarrow -r_3$, respectively).
This is why, in the bilinear formalism, the $Z_2$ and CP1 symmetries are actually denoted $Z_2\times Z_2$ and $Z_2$, respectively.

With arbitrary unitary $2\times 2$ matrices $U$ and $X$ for Higgs-family and GCP field transformations, it would appear that
the number of these symmetries one might impose on the 2HDM potential would be difficult to establish,
but using the bilinear formalism it is simple to see that the maximum number of different such symmetries
is {\it six}~\cite{Ivanov:2005hg,Ivanov:2006yq}. In fact, since in the bilinear formalism symmetry transformations
translate as $O(3)$ rotations imposed on the $\vec{r}$ vector, and any rotation in 3-dimensional space
can be decomposed on parity transformations about the axes, or simple proper rotations about one or more axes,
the total number of different possibilities is:
\begin{itemize}
\item A parity transformation about a single axis. This is the CP1 symmetry, and
the bilinear symmetry group is $Z_2$.
\item A parity transformation about two axes. This is the $Z_2$ symmetry group, and
the bilinear symmetry group is $Z_2\times Z_2$.
\item A parity transformation about the three axes. This is called the CP2 symmetry,
with a bilinear symmetry group $Z_2\times Z_2\times Z_2$. In terms of doublet transformations,
it corresponds to $\Phi_1\rightarrow \Phi_2^*$, $\Phi_2\rightarrow -\Phi_1^*$, but in the
bilinear formalism the corresponding transformation matrix is quite simple:
\beq
S_{CP2}\,=\,\begin{pmatrix} -1 & 0 & 0\\
 0 & -1 & 0 \\
 0 & 0 & -1\end{pmatrix}\, ,
 \label{eq:scp2}
\eeq
\item A rotation about one of the axes. This leads to a $U(1)$ Peccei-Quinn symmetry~\cite{Peccei:1977hh}.
It is obtained requiring invariance under the doublet transformation, $\Phi_1\rightarrow \Phi_1$,
$\Phi_2\rightarrow e^{i\alpha}\Phi_2$ (with $\alpha$ an arbitrary real number), which corresponds to an
$S$ matrix in bilinear space given by
\beq
S_{U(1)}\,=\,\begin{pmatrix} \cos 2\alpha & -\sin 2\alpha & 0\\
\sin 2\alpha & \cos 2\alpha & 0 \\
 0 & 0 & 1\end{pmatrix}\, ,
\eeq
and we recognise a rotation around the third axis, in the plane defined by the first two.
Again, this field/bilinear transformation is expressed in a specific basis, but the potential one would obtain
would be physically equivalent if one were to consider a rotation around any of the other two axes. The symmetry
group in the bilinear formalism is $O(2)$.
\item A rotation about one of the axes along with a parity transformation on the same axis. This is another GCP
symmetry, dubbed CP3, and is obtained via the doublet transformation
\beq
\begin{pmatrix} \Phi_1 \\ \Phi_2\end{pmatrix}\,\rightarrow\,
\begin{pmatrix} \cos\alpha & \sin\alpha \\ -\sin\alpha & \cos\alpha \end{pmatrix}\;
\begin{pmatrix} \Phi_1^* \\ \Phi_2^* \end{pmatrix}\,,
\eeq
where, without loss of generality, $0 <\alpha <\pi/2$. This corresponds to an improper rotation around the
direction of the second axis of $\vec{r}$,
\beq
S_{CP3}\,=\,\begin{pmatrix} \cos 2\alpha & 0 & \sin 2\alpha \\
 0 & -1 & 0 \\
-\sin 2\alpha & 0 & \cos 2 \alpha \end{pmatrix}\, ,
\eeq
corresponding to a $Z_2\times O(2)$ symmetry group in the bilinear formalism.
\item A generic rotation in the 3-dimensional space of the vector $\vec{r}$, corresponding
to the most general matrix $U\in U(2)$ in eq.~\eqref{eq:hf}. This is commonly referred as the $SO(3)$-symmetric
potential, and the rotation matrix in the bilinear formalism is the most generic $SO(3)$ matrix possible.
The bilinear formalism symmetry is therefore $SO(3)$.
\end{itemize}
These are the six symmetries of the $SU(2)\times U(1)$ invariant\footnote{If one disregards hypercharge, the number of symmetries obtained is
larger, including for instance the custodial symmetry group~\cite{Battye:2011jj,Pilaftsis:2011ed}.} 2HDM scalar potential that one can obtain
from unitary or anti-unitary field transformations. In table~\ref{tab:sym} we summarise the impact each symmetry has on the parameters
of the scalar potential. This table considers that each symmetry was imposed in the basis for which the field transformations are
as shown above\footnote{The counting of parameters may seem odd for the CP2 case in the chosen basis. In a simpler basis, proposed
in~\cite{Davidson:2005cw}, the conditions on the model's parameters make $\lambda_5$ real and $\lambda_6 = \lambda_7 = 0$,
yielding 5 independent parameters. Likewise, for the $Z_2$ symmetry, notice that the $\lambda_5$ coupling can always be made real
    through a basis redefinition, which eliminates one of the parameters.}.
\begin{table}[h]
\begin{center}
\begin{tabular}
{|c|ccc|ccccccc|c|}
\hline \hline
 Symmetry &
$m_{11}^2$ & $m_{22}^2$ & $m_{12}^2$ &
$\lambda_1$ & $\lambda_2$ & $\lambda_3$ & $\lambda_4$ &
$\lambda_5$ & $\lambda_6$ & $\lambda_7$ & $N$ \\
\hline
 CP1 &  &  & real &
 & &  &  &
real & real & real & 9 \\
 $Z_2$ &   &   & 0 &   &  &  &  &
    & 0 & 0 & 7 \\
 U(1) &  &  & 0 &
 &  & &  &
0 & 0 & 0 & 6 \\
 CP2 &  & $m_{11}^2$ & 0 &
  & $\lambda_1$ &  &  &
    &   &  -$\lambda_6$ & 5\\
 CP3 &  & $m_{11}^2$ & 0 &
   & $\lambda_1$ &  &  &
$\lambda_1 - \lambda_3 - \lambda_4$  & 0 & 0 & 4 \\
 $SO(3)$ &  & $ m_{11}^2$ & 0 &
   & $\lambda_1$ &  & $\lambda_1 - \lambda_3$ &
0 & 0 & 0 & 3 \\
\hline \hline
\end{tabular}
\end{center}
\caption{Relations between 2HDM scalar potential parameters
 for each of the six symmetries discussed, and the number $N$ of independent real parameters
for each symmetry-constrained scalar potential.
}
\label{tab:sym}
\end{table}

Having reviewed the way the 2HDM global symmetries are obtained, we will argue, in section~\ref{sec:news}
that there are indeed other symmetries not considered in the classification shown above.

\subsection{Renormalization group equations}
\label{sec:rg}

The one-loop renormalization group (RG) equations yield the model's $\beta$-functions. They are given, for
the most general 2HDM of eq.~\eqref{eq:pot}, by~\cite{Haber:1993an,Branco:2011iw,Bednyakov:2018cmx}\footnote{For notational convenience, we suppress a factor $1/(16\pi^2)$.}
\bea
\beta_{m_{11}^2} &=&
3 \lambda_1 m_{11}^2
+ \left( 2 \lambda_3 +  \lambda_4 \right) m_{22}^2
- 3\, \left( \lambda_6^* \,m_{12}^2 + \mathrm{h.c.} \right)
\,-\,\frac{1}{4}\,(9 g^2 + 3 g^{\prime 2}) \,m_{11}^2 \nonumber \\
 & & +\,\beta^F_{m_{11}^2}, \nonumber
\\
\beta_{m_{22}^2} &=&
\left( 2 \lambda_3 +  \lambda_4 \right) \,m_{11}^2
+ 3 \lambda_2 \,m_{22}^2
- 3\, \left( \lambda_7^*\, m_{12}^2+ \mathrm{h.c.} \right)
\,-\,\frac{1}{4}\,(9 g^2 + 3 g^{\prime 2}) \,m_{22}^2
\nonumber \\
 & & +\,\beta^F_{m_{22}^2}, \nonumber
\\
\beta_{m_{12}^2} &=&
- 3 \left( \lambda_6 \,m_{11}^2 + \lambda_7 \, m_{22}^2 \right)
+ \left(  \lambda_3 + 2 \lambda_4 \right) \,m_{12}^2
+ 3 \lambda_5 \,{m_{12}^2}^\ast
\,-\,\frac{1}{4}\,(9 g^2 + 3 g^{\prime 2}) m_{12}^2
\nonumber \\
 & & +\,\beta^F_{m_{12}^2},
\label{eq:betam}
\eea
for the quadratic couplings, and for the quartic ones,
\begin{subequations}
\bea
\beta_{\lambda_1} &=&
6 \lambda_1^2 + 2 \lambda_3^2 + 2 \lambda_3 \lambda_4 +  \lambda_4^2
+  \left| \lambda_5 \right|^2 + 12 \left| \lambda_6 \right|^2 \nonumber \\
 & &
 + \frac{3}{8}(3g^4 + g^{\prime 4} +2 g^2 g^{\prime 2}) -
 \frac{3}{2} \lambda_1 (3 g^2 + g^{\prime 2} )\,+\,\beta^F_{\lambda_1}, \\
%
\beta_{\lambda_2} &=&
6 \lambda_2^2 + 2 \lambda_3^2 + 2 \lambda_3 \lambda_4 +  \lambda_4^2
+  \left| \lambda_5 \right|^2 + 12 \left| \lambda_7 \right|^2 \nonumber \\
 & &
+\
\frac{3}{8}(3g^4 + g^{\prime 4} +2g^2 g^{\prime 2}) -  \frac{3}{2} \lambda_2
(3g^2 +g^{\prime 2})\,+\,\beta^F_{\lambda_2},  \\
%
\beta_{\lambda_3}  &=&
\left( \lambda_1 + \lambda_2 \right) \left( 3 \lambda_3 + \lambda_4 \right)
+ 2 \lambda_3^2 + \lambda_4^2
+  \left| \lambda_5 \right|^2
+ 2 \left( \left| \lambda_6 \right|^2 + \left| \lambda_7 \right|^2 \right)
+ 8\, \mathrm{Re} \left( \lambda_6 \lambda_7^\ast \right) \nonumber \\
 & &
+\ \frac{3}{8}(3g^4 + g^{\prime 4} -2g^2 g^{\prime 2}) - \frac{3}{2}\lambda_3
(3g^2 +g^{\prime 2})\,+\,\beta^F_{\lambda_3}, \\
%
\beta_{\lambda_4}  &=&
 \left( \lambda_1 + \lambda_2 \right) \lambda_4
+ 4 \lambda_3 \lambda_4 + 2 \lambda_4^2
+ 4 \left| \lambda_5 \right|^2
+ 5 \left( \left| \lambda_6 \right|^2 + \left| \lambda_7 \right|^2 \right)
+ 2\, \mathrm{Re} \left( \lambda_6 \lambda_7^\ast \right) \nonumber \\
& &
+ \ \frac{3}{2} g^2 g^{\prime 2} - \frac{3}{2} \lambda_4 (3g^2 +g^{\prime 2})\,+\,\beta^F_{\lambda_4}, \\
%
\beta_{\lambda_5}  &=&
\left( \lambda_1 + \lambda_2 + 4 \lambda_3 + 6 \lambda_4 \right) \lambda_5
+ 5 \left( \lambda_6^2 + \lambda_7^2 \right) + 2 \lambda_6 \lambda_7
\nonumber \\
 & &
-\ \frac{3}{2} \lambda_5 (3g^2 +g^{\prime 2})\,+\,\beta^F_{\lambda_5},  \\
%
\beta_{\lambda_6}  &=&
\left( 6 \lambda_1 + 3 \lambda_3 + 4 \lambda_4 \right) \lambda_6
+ \left( 3 \lambda_3 + 2 \lambda_4 \right) \lambda_7
+ 5 \lambda_5 \lambda_6^\ast +   \lambda_5 \lambda_7^\ast \nonumber \\
 & &
-\ \frac{3}{2} \lambda_6 (3g^2 +g^{\prime 2})\,+\,\beta^F_{\lambda_6},  \\
%
\beta_{\lambda_7}  &=&
\left(6 \lambda_2 +3 \lambda_3 + 4 \lambda_4 \right) \lambda_7
+ \left( 3 \lambda_3 + 2 \lambda_4 \right) \lambda_6
+ 5 \lambda_5 \lambda_7^\ast +  \lambda_5 \lambda_6^\ast \nonumber \\
 & &
-\ \frac{3}{2} \lambda_7 (3g^2 +g^{\prime 2})\,+\,\beta^F_{\lambda_7},
\label{eq:betal}
\eea
\end{subequations}
where the $\beta^F_x$ terms contain all contributions coming from fermions, which we will disregard for the moment, and return
to in section~\ref{sec:ferm}~\footnote{Or we can disregard them altogether and think of the symmetries
existing in a theory without fermions.}.
For simplicity, we have absorbed factors of $16\pi^2$ within the definition of the $\beta$-functions.
$g$ and $g^\prime$, obviously, represent the
SU(2) and $\text{U(1)}$ gauge couplings. The results for the 2HDM two-loop-$\beta$ functions for the quartic couplings
may be found, for instance, in the package
SARAH~\cite{Staub:2008uz,Staub:2009bi,Staub:2010jh,Staub:2012pb,Staub:2013tta}. The 2HDM three-loop
 $\beta$-functions have been obtained by
Bednyakov~\cite{Bednyakov:2018cmx}. The above $\beta$-functions allow us to verify that the
relations obtained in the previous section among parameters are RG-invariant to one-loop order. For
instance, we observe that if all of the quartic
couplings are real, as consequence of a CP1 symmetry, no imaginary components for the $\lambda_i$ are
generated at one-loop. Likewise, we see that if
one imposes a $Z_2$ symmetry so that $\lambda_6 = \lambda_7 = 0$ one immediately obtains
$\beta_{\lambda_6} = \beta_{\lambda_7} = 0$, confirming
that the symmetry-obtained condition on the $\lambda$'s is preserved under radiative corrections
at the one-loop order. Indeed, we may expect that
condition to hold to all orders of perturbation theory, precisely because it is obtained via a
symmetry. Another interesting perspective is
obtained looking at the $\lambda_5$ $\beta$-function for the $Z_2$ model,
\beq
\beta_{\lambda_5}  \,=\,
\left[ \lambda_1 + \lambda_2 + 4 \lambda_3 + 6 \lambda_4 - \ \frac{3}{2} \ (3g^2 +g^{\prime 2})\right] \,\lambda_5
\,,
\eeq
wherein one identifies a {\em fixed point} of this RG equation -- if at any scale one should have $\lambda_5 = 0$, that coupling
will remain equal to zero for all renormalization scales. Such fixed points of RG equations are usually fingerprints of hidden
symmetries, and indeed that is the case here: if $\lambda_6 = \lambda_7 = 0$, the extra constraint $\lambda_5 = 0$ takes us from a
$Z_2$-symmetric model to a $U(1)$-symmetric one, as can be seen from table~\ref{tab:sym}.

At this point, and as it will be crucial for the discussion in the next section, let us observe that the set of conditions
\beq
\left\{m^2_{11} + m^2_{22} = 0\;\;,\;\; \lambda_1 = \lambda_2 \;\;,\;\; \lambda_6 = -\lambda_7\right\}
\label{eq:nsc}
\eeq
constitutes a fixed point of the one-loop RG equations.  In fact, by manipulating the above
$\beta$-functions, we obtain
\bea
\beta_{m_{11}^2 + m_{22}^2} &=&
3 (\lambda_1 m_{11}^2 + \lambda_2 m_{22}^2)
+ \left( 2 \lambda_3 + \lambda_4 \right) (m_{11}^2 + m_{22}^2)
- 3\, \left[ (\lambda_6^\ast + \lambda_7^\ast) m_{12}^2 + \mathrm{h.c.} \right]\nonumber \\
& & -\,\frac{1}{4}\,(9 g^2 + 3 g^{\prime 2}) (m_{11}^2 + m_{22}^2)\,
,
\\
\beta_{\lambda_1-\lambda_2} &=&
6 \left(\lambda_1^2 - \lambda_2^2\right) + 12 \left(\left| \lambda_6 \right|^2 -
\left| \lambda_7 \right|^2\right)
- \frac{3}{2} (\lambda_1 - \lambda_2)\,(3 g^2 + g^{\prime 2} ), \\
\beta_{\lambda_6 + \lambda_7 }  &=& 6\left(\lambda_1 \lambda_6 + \lambda_2 \lambda_7\right) +
\left( 3 \lambda_3 + 2 \lambda_4 \right) \left(\lambda_6 + \lambda_7\right)
+ 6 \lambda_5 \left(\lambda_6^\ast + \lambda_7^\ast \right) \nonumber \\
 & & -\ \frac{3}{2}(\lambda_6 + \lambda_7)\,(3g^2 +g^{\prime 2})\,,
\eea
and we see that the conditions listed in eq.~\eqref{eq:nsc} do constitute a fixed point of these RG equations.
Of course, it must be said that just because the one-loop $\beta$-functions have a fixed point that is not
guaranteed to indicate a symmetry -- it may be, unlike the $U(1)$ example discussed above, simply a one-loop
accident that such a fixed point occurs. As we will argue in the next section, though, that is not the case,
and the conditions of eq.~\eqref{eq:nsc} are indeed invariant for all orders of perturbation theory.

We also take the opportunity to point out that the parameter conditions of eq.~\eqref{eq:nsc} are {\em basis
invariant}. This can be shown explicitly by using the general basis transformations presented in
eqs.~\eqref{eq:M11}--\eqref{eq:L7}.

Finally, we remark that the two relations between quartic couplings in eq.~\eqref{eq:nsc} may look familiar:
they are exactly the ones we obtain from the application of the CP2 symmetry (check table~\ref{tab:sym}).
Notice, however, that the conditions on the quadratic parameters arising from CP2 are {\em not} the same
as those in eq.~\eqref{eq:nsc}. We will return to this subject shortly.

\section{New 2HDM symmetries}
\label{sec:news}

In this section we will argue that new symmetries of the 2HDM $SU(2)\times U(1)$ scalar potential of eq.~\eqref{eq:pot}
exist, other than those discussed in section~\ref{sec:sym}. We will arrive at this conclusion by identifying all-order
fixed points in the 2HDM $\beta$-functions, and to reach that argument we will use a curious interplay between basis
invariance, dimensional analysis and RG equations.

\subsection{All-orders fixed points in 2HDM RG equations}

As explained in section~\ref{sec:basis}, basis transformations are extremely simple to represent in the bilinear formalism.
A generic basis transformation corresponds, in bilinear space, to a generic $O(3)$ rotation matrix $R$, and as such
$\vec{r}$, $\vec{M}$ and $\vec{\Lambda}$ transform as 3-vectors under these rotations; the $3\times 3$ matrix $\Lambda$
is also transformed as under a rotation in this space; and the quantities $r_0$, $M_0$ and $\Lambda_{00}$ are basis invariants.
It is then possible to write the most generic set of basis invariant quantities one can form with the quartic parameters of the
potential~\cite{Ivanov:2005hg,Bednyakov:2018cmx}. These are
\begin{align}
I_{1,1}&=\,\Lambda_{00}\,, & I_{1,2} &=\, \Tr\Lambda \nonumber \\
I_{2,1}&=\,\vec{\Lambda} \cdot\vec{\Lambda}  \,, & I_{2,2} &=\, \Tr\Lambda^2 \nonumber \\
I_{3,1}&=\,\vec{\Lambda} \cdot\Lambda \vec{\Lambda}  \,, & I_{3,2}&=\, \Tr\Lambda^3 \nonumber \\
I_{4,1}&=\,\vec{\Lambda} \cdot\Lambda^2  \vec{\Lambda}  \,, &  &
\label{eq:inv4}
\end{align}
One might think that higher powers of the $\Lambda$ matrix could be used to build further invariants,
but that is not the case. In fact, this matrix satisfies~\cite{Bednyakov:2018cmx}
\beq
\Lambda^3\,=\, (\Tr\Lambda)\Lambda^2
-\frac{1}{2} \left[(\Tr\Lambda)^2 - \Tr\Lambda^2\right] \Lambda
+\frac{1}{6} \left[(\Tr\Lambda)^3 -3 \Tr\Lambda\,\Tr\Lambda^2 +
2\Tr\Lambda^3\right]\,\mathbb{1}_{3\times 3}\,.
\label{eq:lam3}
\eeq
This relation, obtained via
the Cayley-Hamilton theorem, shows that powers of $\Lambda$ higher than 2 can always be expressed as
a sum of powers of up to 2 of that matrix~\footnote{This is also the reason why we do not need to consider
the basis-invariant determinant of $\Lambda$ in this discussion, since that the determinant of a
$3\times 3$ matrix can be expressed as a linear combination of the traces of its powers up to 3.}.

As explained in~\cite{Bednyakov:2018cmx}, then, the $\beta$ function of the vector $\vec{\Lambda}$ is given,
to all orders of perturbation theory, by
\beq
\beta_{\vec{\Lambda}}\,=\, a_0 \,\vec{\Lambda} \,+\, a_1\, \Lambda\, \vec{\Lambda}\, +\,
a_2\, \Lambda^2\, \vec{\Lambda}
\label{eq:betaLv}
\eeq
where the $a_i$ are polynomial expressions involving the invariants of eq.~\eqref{eq:inv4}. If one computes
this $\beta$-function at an arbitrary number of loops in perturbation theory, basis invariance will always
require that it is given by the structure shown above. Indeed, eq.~\eqref{eq:betaLv}
expresses a very elegant interplay between basis invariance and RG equations: since $\vec{\Lambda}$ transforms
as a vector under basis transformations, its $\beta$-function must transform in the same manner; therefore,
the right-hand side of~\eqref{eq:betaLv} must be composed of terms proportional to vector-like combinations of
couplings, and the only three that can be used are $\vec{\Lambda}$,  $\Lambda \,\vec{\Lambda}$ and $\Lambda^2 \,\vec{\Lambda}$
-- higher powers of $\Lambda$, as explained above, are superfluous. There is another vector for basis transformation
involving couplings of the potential -- $\vec{M}$ -- but due to its dimensions of mass, it cannot enter in~\eqref{eq:betaLv}.
This argument can easily be extended to accommodate the contributions from the gauge couplings -- as the gauge sector
is left unchanged under basis transformations, terms involving the couplings $g$ and $g^\prime$ will simply contribute
to the coefficients $a_i$ in~\eqref{eq:betaLv}.

With basis transformation properties dictating that the structure of $\beta_{\vec{\Lambda}}$ is, to all orders, a series of terms all
linear in $\vec{\Lambda}$, we reach a straightforward conclusion:
\begin{itemize}
\item{\em $\vec{\Lambda} = \vec{0}$ is a fixed point of the RG equation for
this quantity, to all orders of perturbation theory}.
\end{itemize}
Now, $\vec{\Lambda} = \vec{0}$ implies, in terms of the notation of eq.~\eqref{eq:pot},
that $\lambda_1 = \lambda_2$ and $\lambda_6 = -\lambda_7$, which are the conditions on quartic couplings we discussed in eq.~\eqref{eq:nsc}.
They are also, as we already mentioned, the conditions one obtains for the quartic couplings from the CP2 symmetry.
So this $\beta$-function argument seems to have led us to re-discover the CP2 symmetry, but as we will shortly see that is not
necessarily so.

Continuing to follow the reasoning of~\cite{Bednyakov:2018cmx}, the $\beta$-function for the quadratic parameter
singlet $M_0 = m^2_{11} + m^2_{22}$ defined in eq.~\eqref{eq:defbil} must obey two constraints: it must have dimensions
of (mass)$^2$
and it must be a singlet under basis transformations. Given the property of the $\Lambda$ matrix shown
in eq.~\eqref{eq:lam3}, it is easy to conclude that $\beta_{M_0}$ is a linear combination, via basis invariant
dimensionless coefficients $b_i$, of four different quantities,
\beq
\beta_{M_0}\,=\, b_0 \,M_0\,+\,b_1\,\vec{\Lambda} \cdot\vec{M}\,+\, b_2 \,\vec{\Lambda} \cdot\left(\Lambda \vec{M}\right)
\,+\, b_3 \,\vec{\Lambda} \cdot\left(\Lambda^2 \vec{M}\right)\,.
\label{eq:betaM0}
\eeq
It is easy to understand the structure of this equation -- since all terms must have the same mass dimension they are either
built with $M_0$ or the vector $\vec{M}$; and any term involving $\vec{M}$ must involve an internal product with a dimensionless
vector to form a basis transformation singlet, and the only such vector available is $\vec{\Lambda}$. And as before, this structure
is easily generalizable to include gauge couplings -- since there are no other terms in the 2HDM lagrangian with the appropriate dimensions,
all gauge contributions will simply be contained in the $b_i$ coefficients of eq.~\eqref{eq:betaM0}. The structure of this equation
also allows us to reach a simple conclusion:
\begin{itemize}
\item{\em If $\vec{\Lambda} = \vec{0}$, then $M_0 = 0$ is a fixed point of the RG equation for
this quantity, to all orders of perturbation theory}.
\end{itemize}
Following the same line of reasoning, the $\beta$-function for the vector $\vec{M}$ of eq.~\eqref{eq:defbil} should be given by
a linear combination of terms with dimension (mass)$^2$ which behave as vectors under basis transformations. This leads us to
\beq
\beta_{\vec{M}} \,=\, c_0 \,\vec{M}\,+\,c_1\,\Lambda\, \vec{M}\,+\,c_2\,\Lambda^2\,\vec{M}\,+\,c_3\,I_M\,\vec{\Lambda}\,+\,
c_4 \, I_M\,\Lambda\, \vec{\Lambda}\,+\,c_5\,I_M\,\Lambda^2\,\vec{\Lambda}\,
\eeq
where $I_M$ stands for some linear combination of the four basis-invariant quantities (with the same dimension as $\vec{M}$)
used in eq.~\eqref{eq:betaM0}. And once again, we see that this RG equation possesses a fixed point:
\begin{itemize}
\item{\em If $\vec{\Lambda} = \vec{0}$, then $\vec{M} = \vec{0}$ is a fixed point of the RG equation for
this quantity, to all orders of perturbation theory.}
\end{itemize}
Notice how the existence of this fixed point is completely independent of the previous one. We have therefore identified
two all-orders fixed points of the 2HDM RG equations:
\begin{itemize}
\item {\em $\{\vec{M} = \vec{0}\,,\,\vec{\Lambda} = \vec{0}\}$. This is equivalent, in the notation of~\eqref{eq:pot}, to}
\beq
m^2_{11}\,=\,m^2_{22}\;\;,\;\;m^2_{12} = 0\;\;\; , \;\;\; \lambda_1 = \lambda_2 \;\;,\;\; \lambda_6 = -\lambda_7\,.
\eeq
These are exactly the CP2 symmetry conditions.
\item  {\em $\{M_0 = 0\,,\,\vec{\Lambda} = \vec{0}\}$. This is equivalent, in the notation of~\eqref{eq:pot}, to}
\beq
m^2_{11}\,=\,-m^2_{22}\;\;\; , \;\;\; \lambda_1 = \lambda_2 \;\;,\;\; \lambda_6 = -\lambda_7\,.
\eeq
These are the conditions mentioned before in eq.~\eqref{eq:nsc} and they coincide with the CP2 symmetry conditions for
the quartic couplings, but have different conditions for the
quadratic ones. As we have already discussed these conditions are basis invariant, so they are {\em not} a basis
change of the previous ones.
\end{itemize}
We have already shown explicitly that the conditions $\{M_0 = 0\,,\,\vec{\Lambda} = \vec{0}\}$ ({\em i.e} eq.~\eqref{eq:nsc})
are RG-invariant at the one-loop level. The reader is encouraged to verify, as we have done, that that statement holds at least to two
and three-loop level, using the explicit results for the $\beta$-functions of~\cite{Staub:2008uz,Bednyakov:2018cmx}.

It may be tempting to think of the above second set of conditions on the parameters of the potential as a special soft breaking
version of the CP2 model. In fact, it is not unheard of that some soft breaking conditions are RG invariant. We can imagine
one such scenario for the CP2 model -- according to table~\ref{tab:sym}, the exact CP2 symmetry implies $m^2_{11} = m^2_{22}$
and $m^2_{12} = 0$. If one now considers a softly broken potential with $m^2_{12} \neq 0$, the condition $m^2_{11} = m^2_{22}$
will be RG-preserved to all orders, since this potential has a residual $S_2$ permutation symmetry ($\Phi_1\leftrightarrow \Phi_2$).
If instead one were to consider  a softly broken potential with $m^2_{11} \neq m^2_{22}$ one would still have $m^2_{12} = 0$
at all orders of perturbation theory, since this model has a residual $Z_2$ symmetry.

However, notice that if the set of constraints $\{M_0 = 0\,,\,\vec{\Lambda} = \vec{0}\}$ is satisfied,
that imposes conditions on the quadratic part of the potential which are
(a)  invariant to all orders of perturbation theory and (b) different from any conditions any of the six
symmetries listed in table~\ref{tab:sym} manages to impose on those parameters. In fact, the most that Higgs-family
or GCP symmetries manage to do about the quadratic parameters is impose the equality of $m^2_{11}$ and $m^2_{22}$,
the realness of $m^2_{12}$ or its vanishing -- {\em never} such a distinct relation as $m^2_{11} = -m^2_{22}$.
Indeed, this all-order constraint imposed on the quadratic parameters cannot be obtained via the two types of symmetries
we have been discussing -- how then can we obtain it? In the following section we will provide a simple interpretation,
in the bilinear formalism, of how $\{M_0 = 0\,,\,\vec{\Lambda} = \vec{0}\}$ may arise, and argue it constitutes a new type of
2HDM symmetry.

\subsection{The $r_0$ symmetry - bilinear interpretation}
\label{sec:r0}

Let us begin by remarking that another useful way of writing the scalar potential of
eq.~\eqref{eq:pot} is by making obvious the dependence on the basis invariant and vector-like objects. This is
very easily expressed in terms of the bilinear formalism and the quantities defined in
eqs.~\eqref{eq:defbil}--\eqref{eq:L0Lv}, so that
\beq
V \,=\, M_0\,r_0\,+\,\Lambda_{00}\,r_0^2 \,-\,{\vec{M}}\,. \, \vec{r}
\,-\,2\,\left({\vec{\Lambda}}\,.\,\vec{r}\right)\,r_0 \,+\,
 {\vec{r}\,}\,.\,\left(\Lambda\,\vec{r}\right)\,.
 \label{eq:potbi}
\eeq

As defined in section~\eqref{sec:sym}, the CP2 symmetry corresponds, in the bilinear formalism, to a parity transformation
about the three axes of the vector $\vec{r}$, such that $\vec{r}\rightarrow - \vec{r}$. Applied to the potential written
in the bilinear notation above, it is immediate to see what the result of the application of CP2 is: the potential can only
remain invariant under the symmetry if $\vec{\Lambda} = \vec{0}$ and $\vec{M} = \vec{0}$.

The bilinear writing of the potential makes it also clear that there is seemingly {\em another} way to obtain $\vec{\Lambda} = \vec{0}$.
To  wit, consider what happens to the potential if one changes the sign of $r_0$:
\beq
r_0 \rightarrow - r_0\; \Longrightarrow \; \{M_0 = 0 \; , \;  \vec{\Lambda} = \vec{0}\}.
\eeq
These are exactly the conditions we obtained from the second all-orders fixed point identified above, that lead to the parameter relations
shown in eq.~\eqref{eq:nsc}. Let us call this the {\em $r_0$ symmetry}.

The seminal work of~\cite{Ivanov:2006yq,Ivanov:2007de} did not consider any transformations of the type $r_0 \rightarrow - r_0$, for two
very good reasons: first, the way $r_0$ is defined (check eq.~\eqref{eq:rs}), this quantity is always positive; second, $r_0$ is
left invariant under any unitary or anti-unitary doublet transformations, which compose both Higgs-family and GCP symmetries. The first
of these objections can be remedied: eq.~\eqref{eq:rs} can be trivially changed, so that $r_0$ is defined as having both signs,
with a simultaneous change in the signs of the $r_i$:
\beq
\begin{array}{rcl}
r_0 &=&
\pm\,\frac{1}{2}
\left( \Phi_1^\dagger \Phi_1 + \Phi_2^\dagger \Phi_2 \right),
\\*[2mm]
r_1 &=&
\pm\,\frac{1}{2}
\left( \Phi_1^\dagger \Phi_2 + \Phi_2^\dagger \Phi_1 \right)
= \pm\mbox{Re}\left( \Phi_1^\dagger \Phi_2 \right),
\\*[2mm]
r_2 &=&
\mp\,\frac{i}{2}
\left( \Phi_1^\dagger \Phi_2 - \Phi_2^\dagger \Phi_1 \right)
= \pm\mbox{Im} \left( \Phi_1^\dagger \Phi_2 \right),
\\*[2mm]
r_3 &=&
\pm\,\frac{1}{2}
\left( \Phi_1^\dagger \Phi_1 - \Phi_2^\dagger \Phi_2 \right).
\end{array}
\eeq
By expanding the range of variation of $r_0$ we should in principle cause no changes in the conclusions
derived in~\cite{Ivanov:2006yq,Ivanov:2007de} for the positivity conditions on the potential, or the number
and types of minima possible, since we simultaneously force a change in the sign of the $r_i$.

As for the second consideration, it is part of the reason why we argue that the conditions of eq.~\eqref{eq:nsc}
constitute a new type of 2HDM symmetry -- we have shown that they are preserved under renormalization to all
orders of perturbation theory, which is the hallmark of the presence of a symmetry. We have shown that they
can be obtained, at least formally, via a parity transformation on the ``time" axis of the $r_\mu$ bilinear vector.
There is no unitary or antiunitary doublet transformation that can yield $r_0 \rightarrow - r_0$, nor can such
transformations yield a parameter condition like $M_0 = 0 \Longleftrightarrow m^2_{11} = - m^2_{22}$. Nonetheless,
that condition was found to be both basis invariant {\em and} RG invariant to all orders. The six symmetries
described in section~\ref{sec:sym} can be described via transformations on the doublets, which have a
counterpart as transformations on the bilinears -- for the $r_0$ symmetry, we can obtain RG-invariant conditions
on the potential via a bilinear transformation, which seemingly has no equivalent on transformations expressed
in terms of the doublets themselves. In this regard, it is almost as if the bilinear formalism is more ``fundamental"
in what concerns the scalar sector of the 2HDM, as aspects of the model can be understood in terms of the $r_\mu$
but not in terms of the $\Phi_i$.

We must worry about the kinetic terms too, however, and in particular the gauge interactions of the doublets.
Here, again, the limitations of the bilinear formalism make themselves manifest.
In eq.~\eqref{eq:kin} the 2HDM kinetic terms were written using the same Minkowski formalism used for the
bilinears and the potential, but not considering the gauge interactions. The doublets' covariant
derivatives are defined as
\beq
D_\mu\,=\,\partial_\mu\,+\,i \frac{g^\prime}{2}Y B_\mu \,+\,i \frac{g}{2}\sigma^a W^a_\mu\,,
\eeq
where $Y$ is the hypercharge of the fields the derivative operates upon, an implicit sum on $a = 1, 2, 3$
is assumed and $W^a_\mu$ and $B_\mu$ are the $SU(2)_L$ and $U(1)_Y$ gauge fields respectively. The kinetic terms are therefore
given by (using the fact that both scalar doublets have hypercharge $Y = 1$)
\bea
T  &=& 
\left(D_\mu \Phi_i\right)^\dagger\,D^\mu \Phi_i \nonumber \\
 &=& 
  \partial_\mu \Phi_i^\dagger\,\partial^\mu \Phi_i\,+\,
 \frac{i g^\prime}{2}\,\left[\left(\partial_\mu \Phi_i^\dagger\right)\,\Phi_i \,-\, \Phi_i^\dagger \,\left(\partial_\mu \Phi_i\right)\right]\,B^\mu \,+\,
 \frac{i g}{2}\,\left[\left(\partial_\mu \Phi_i^\dagger\right)\sigma^a\Phi_i \,-\, \Phi_i^\dagger \sigma^a \left(\partial^\mu \Phi_i\right)\right]\,
 W^{a \mu}   \nonumber \\
 & &  +\, \frac{1}{2}\,g\,g^\prime\,\left(\Phi_1^\dagger\sigma^a\Phi_1 + \Phi_2^\dagger\sigma^a\Phi_2\right)\,W^a_\mu B^\mu
 \,+\,\frac{1}{4}\,\left({g^\prime}^2\,B_\mu B^\mu + g^2\,W^a_\mu W^{a\mu}\right)\,\left(|\Phi_1|^2 + |\Phi_2|^2 \right)\,,
\eea
where again an implicit sum on the indices $i = 1, 2$ and $a = 1, 2, 3$ is assumed.
Hence, we can rewrite this equation as
\bea
T  &=& K_1^{\mu}\,\left\{ \left(\partial_\alpha \Phi_i^\dagger\right) \,(\sigma_\mu)_{ij}\,
\left(\partial^\alpha \Phi_j\right)\,+\,
\frac{i g^\prime}{2}\,\left[\left(\partial_\alpha \Phi_i^\dagger\right)\,(\sigma_\mu)_{ij}\,\Phi_j
\,-\, \Phi_i^\dagger \,(\sigma_\mu)_{ij}\,\left(\partial_\alpha \Phi_j\right)\right]\,B^\alpha  \right.
\nonumber \\
& & \quad\quad \left. +\, \frac{i g}{2}\,\left[\left(\partial_\alpha \Phi_i^\dagger\right)
\,(\sigma_\mu)_{ij}\, \sigma^a\Phi_j \,-\, \Phi_i^\dagger \,(\sigma_\mu)_{ij}\, \sigma^a
\left(\partial^\alpha \Phi_j\right)\right]\, W^{a \alpha} \,+\,
\frac{g\,g^\prime}{2}\, \Phi_i^\dagger\,(\sigma_\mu)_{ij}\,\sigma^a\,\Phi_j\, W^a_\alpha B^\alpha\right\}
  \nonumber \\
  & &
 \, +\,\frac{1}{2}\,K_2^\mu \,\left({g^\prime}^2\,B_\alpha B^\alpha + g^2\,W^a_\alpha W^{a\alpha}\right)\,r_\mu\,,
\eea
with $K_1^\mu \,=\,K_2^\mu \,=\,(1\,,\,0\,,\,0\,,\,0)$, and care must be taken to not confuse the 4-vector $\sigma_\mu$
(defined in eq.~\eqref{eq:kin})
and the three Pauli matrices $\sigma^a$. The last term can be made to remain
 invariant under the transformation $r_0\rightarrow -r_0$ if one assumes $K_2\rightarrow -K_2$, as well. However,
 that does not explain how the remaining terms, involving derivatives and gauge fields, could remain invariant.
This once more emphasizes that we do not know what the expression of the
$r_0$ symmetry in terms of doublet fields (and their derivatives) ought to be. However, in appendix~\ref{ap:comp}
we show how a peculiar transformation of fields and spacetime coordinates could reproduce the $r_0$
symmetry, at least formally.

However, though we may be unable
to write the kinetic terms in a satisfactory way as a function of bilinears, this does not invalidate the
fact that the region of parameter space we identify with the $r_0$ symmetry is RG invariant to all orders,
and we must remember that that reasoning included the contributions of gauge interactions as well.

We therefore argue that the conditions of eq.~\eqref{eq:nsc}, which are basis and RG invariant, are obtained
from the imposition on the potential of a new type of symmetry, which we have dubbed the $r_0$ symmetry. We
have provided a bilinear transformation which, applied to the potential, yields these conditions on the
parameters of the potential. Though the conditions on the quartic couplings can be obtained via a GCP
symmetry (CP2), no unitary or antiunitary field transformations can reproduce the all-orders
RG-invariant conditions on the quadratic parameters of eq.~\eqref{eq:nsc}. Of course, there are plenty of examples
of symmetries in particle physics models which do not involve this type of transformations, such as supersymmetry,
for instance.

\subsection{List of new symmetries}
\label{sec:newsy}

The $r_0$ symmetry yields CP2-like quartic couplings and $m^2_{11} = - m^2_{22}$. When combined with the bilinear transformations
which yield the six symmetries listed on table~\ref{tab:sym}, we can obtain a total of seven new symmetry classes. We will
designate the new symmetries with the prefix ``0" -- so for instance, ``0CP1" will refer to the application of the $r_0$ and CP1
symmetries, as ``0$Z_2$" refers to the application of $r_0$ and $Z_2$. We therefore obtain the constraints on the parameters of
the potential shown in table~\ref{tab:nsym}.

\begin{table}[h]
\begin{center}
\begin{tabular}
{|c|ccc|ccccccc|}
\hline \hline
 Symmetry &
$m_{11}^2$ & $m_{22}^2$ & $m_{12}^2$ &
$\lambda_1$ & $\lambda_2$ & $\lambda_3$ & $\lambda_4$ &
$\lambda_5$ & $\lambda_6$ & $\lambda_7$ \\
\hline
 $r_0$ &  & $-m^2_{11}$ & &
 & $\lambda_1$ & &  &
  &   & $-\lambda_6$  \\
 0CP1 &  & $-m^2_{11}$ & real &
 & $\lambda_1$ & &  &
real & real & $-\lambda_6$  \\
 0$Z_2$ &   & $-m^2_{11}$  & 0 &  & $\lambda_1$ &  &  &
   & 0 & 0  \\
 0U(1) &   & $-m^2_{11}$  & 0 &  & $\lambda_1$ &  &  &
 0  & 0 & 0 \\
 0CP2 & 0 & 0 & 0 &
  & $\lambda_1$ &  &  &
    &   &  $-\lambda_6$  \\
 0CP3 & 0 & 0 & 0 &
   & $\lambda_1$ &  &  &
$\lambda_1 - \lambda_3 - \lambda_4$  & 0 & 0 \\
 0$SO(3)$ & 0 & 0 & 0 &
   & $\lambda_1$ &  & $\lambda_1 - \lambda_3$ &
0 & 0 & 0  \\
\hline \hline
\end{tabular}
\end{center}
\caption{Relations between 2HDM scalar potential parameters
 for each of the new seven symmetries discussed.
}
\label{tab:nsym}
\end{table}

The last three symmetries listed in table~\ref{tab:nsym} have the odd property of not having any quadratic
parameters -- the combination of the $r_0$ symmetry with others eliminating all of those coefficients. We reached
the parameter relation $m^2_{22} = -m^2_{11}$ through an analysis of all-orders RG invariance, and of course that,
due to dimensional analysis, for any potential with all quadratic couplings vanishing they will remain
zero at all orders of perturbation theory. Such models, however, are clearly not interesting, since electroweak symmetry
breaking is not possible with vanishing quadratic couplings\footnote{Though it might occur when radiative
corrections are taken into account, as in the Coleman-Weinberg mechanism~\cite{Coleman:1973jx}.}. However, soft breaking versions of such models, in particular soft
breakings which include the condition $m^2_{22} = -m^2_{11}$, may be of
interest, and we will consider several such cases in section~\ref{sec:ferm}.

The parameter relations presented in table~\ref{tab:nsym} are not in the simplest form that the 2HDM potential can
have under each of those symmetries, since basis freedom can still be used to eliminate some spurious parameters.
In particular, we can use the result of refs.~\cite{Gunion:2005ja,Davidson:2005cw}, in which it was shown that if
$\lambda_1 = \lambda_2$ and $\lambda_7 = - \lambda_6$, then a basis exists for which all $\lambda_i$ are real and
$\lambda_6 = \lambda_7 = 0$, without any loss of generality. Proceeding to this basis, we obtain the most simple
form of the potential for each symmetry, and can establish the number of independent parameters for each case.
We list the relations between couplings in this new basis, and the number $N$ of free parameters, in table~\ref{tab:nsyms}.
Some of the symmetries shown in table~\ref{tab:nsym} already had $\lambda_6 = \lambda_7 = 0$, so for those there is no change.

\begin{table}[hbt]
\begin{center}
\begin{tabular}
{|c|ccc|ccccccc|c|}
\hline \hline
 Symmetry &
$m_{11}^2$ & $m_{22}^2$ & $m_{12}^2$ &
$\lambda_1$ & $\lambda_2$ & $\lambda_3$ & $\lambda_4$ &
$\lambda_5$ & $\lambda_6$ & $\lambda_7$ & $N$ \\
\hline
 $r_0$ &  & $-m^2_{11}$ & &
 & $\lambda_1$ & &  & real
  & 0 & 0 & 7 \\
 0CP1 &  & $-m^2_{11}$ & real &
 & $\lambda_1$ & &  &
real & 0 & 0 & 6 \\
 0$Z_2$ &   & $-m^2_{11}$  & 0 &  & $\lambda_1$ &  &  &
   real & 0 & 0 & 5 \\
 0U(1) &   & $-m^2_{11}$  & 0 &  & $\lambda_1$ &  &  &
 0  & 0 & 0 & 4 \\
 0CP2 & 0 & 0 & 0 &
  & $\lambda_1$ &  &  & real
    & 0 &  0 & 4 \\
 0CP3 & 0 & 0 & 0 &
   & $\lambda_1$ &  &  &
$\lambda_1 - \lambda_3 - \lambda_4$  & 0 & 0 & 3\\
 0$SO(3)$ & 0 & 0 & 0 &
   & $\lambda_1$ &  & $\lambda_1 - \lambda_3$ &
0 & 0 & 0 & 2 \\
\hline \hline
\end{tabular}
\end{center}
\caption{Relations between 2HDM scalar potential parameters
 for each of the new symmetries in a special basis, and the number $N$ of independent real parameters
for each symmetry-constrained scalar potential.
}
\label{tab:nsyms}
\end{table}

Again, any soft breaking of the above potentials preserves the renormalizability of the model, in particular
the relations between the quartic couplings.

\section{Scalar phenomenology of the new symmetric models}
\label{sec:sphen}

We have shown how the condition $m^2_{22} = -m^2_{11}$, coupled with $\lambda_1 = \lambda_2$ and $\lambda_7 = -\lambda_6$,
constitutes an all-orders RG invariant region of parameter space, which can seemingly be obtained in the bilinear formalism
via the transformation $r_0\rightarrow -r_0$. We now wish to investigate the consequences that this condition in particular
can have on the phenomenology of the 2HDM scalars. To do this we must investigate how electroweak symmetry breaking occurs. For this purpose, we start by writing out our potential in a basis in which the $r_0$ symmetry is manifest. It is given by
\bea
V &=& m_{11}^2\left[\Phi_1^\dagger\Phi_1-\Phi_2^\dagger\Phi_2\right]
-\left[m_{12}^2\Phi_1^\dagger\Phi_2+{\rm h.c.}\right]+\half\lambda_1\left[(\Phi_1^\dagger\Phi_1)^2
+(\Phi_2^\dagger\Phi_2)^2\right]
+\lambda_3(\Phi_1^\dagger\Phi_1)(\Phi_2^\dagger\Phi_2)\nonumber\\[8pt]
&&\quad
+\lambda_4(\Phi_1^\dagger\Phi_2)(\Phi_2^\dagger\Phi_1)
+\left\{\half\lambda_5(\Phi_1^\dagger\Phi_2)^2
+\lambda_6\left[(\Phi_1^\dagger\Phi_1)
-(\Phi_2^\dagger\Phi_2)\right]
\Phi_1^\dagger\Phi_2+{\rm h.c.}\right\}\,,
\label{eq:potr0}
\eea
where all parameters are real, except for $m_{12}^2$, $\lambda_5$ and $\lambda_6$ which may be complex.
Without loss of generality one can rotate into a simpler basis in which $\lambda_6=\lambda_7=0$ and $\lambda_5$ is real to get
\bea
V &=& m_{11}^2\left[\Phi_1^\dagger\Phi_1-\Phi_2^\dagger\Phi_2\right]
-\left[m_{12}^2\Phi_1^\dagger\Phi_2+{\rm h.c.}\right]+\half\lambda_1\left[(\Phi_1^\dagger\Phi_1)^2
+(\Phi_2^\dagger\Phi_2)^2\right]
+\lambda_3(\Phi_1^\dagger\Phi_1)(\Phi_2^\dagger\Phi_2)\nonumber\\[8pt]
&&\quad
+\lambda_4(\Phi_1^\dagger\Phi_2)(\Phi_2^\dagger\Phi_1)
+\frac{\lambda_5}{2}\left[(\Phi_1^\dagger\Phi_2)^2
+(\Phi_2^\dagger\Phi_1)^2\right]\,.
\label{eq:potr0red}
\eea
The Higgs doublets can be parameterized as
\begin{equation}
	\Phi_j=e^{i\xi_j}\left(
	\begin{array}{c}\varphi_j^+\\ (v_j+\eta_j+i\chi_j)/\sqrt{2}
	\end{array}\right), \quad
	j=1,2.\label{vevs}
\end{equation}
Here $v_j$ are real numbers, so that $v_1^2+v_2^2=v^2$. The fields $\eta_j$ and $\chi_j$ are real, whereas $\varphi_j^+$ are complex fields.
Then the most general form of the vacuum will have the form
\bea
\left<\Phi_j\right>=\frac{e^{i\xi_j}}{\sqrt{2}}\left(
\begin{array}{c}0\\ v_j
\end{array}\right),\quad j=1,2,
\eea
where we may without loss of generality choose $\xi_1=0$ and put $\xi_2\equiv \xi$. We may also assume that both $v_i\geq0$.
Massless Goldstone states are extracted by defining  orthogonal states
\beq
\left(
\begin{array}{c}G_0\\ \eta_3
\end{array}\right)
=
\left(
\begin{array}{cc}v_1/v & v_2/v\\ -v_2/v & v_1/v
\end{array}\right)
\left(
\begin{array}{c}\chi_1\\ \chi_2
\end{array}\right),\quad
\left(
\begin{array}{c}G^\pm\\ H^\pm
\end{array}\right)
=
\left(
\begin{array}{cc}v_1/v & v_2/v\\ -v_2/v & v_1/v
\end{array}\right)
\left(
\begin{array}{c}\varphi_1^\pm\\ \varphi_2^\pm
\end{array}\right).
\eeq
Then $G_0$ and $G^\pm$ become the massless Goldstone fields, and $H^\pm$ are the charged scalars.
The model also contains three neutral scalars, which are linear compositions of the $\eta_i$,
\begin{equation} \label{Eq:R-def}
	\begin{pmatrix}
		H_1 \\ H_2 \\ H_3
	\end{pmatrix}
	=R
	\begin{pmatrix}
		\eta_1 \\ \eta_2 \\ \eta_3
	\end{pmatrix},
\end{equation}
with the $3\times3$ orthogonal rotation matrix $R$ satisfying
\begin{equation}
	R{\cal M}^2R^{\rm T}={\cal M}^2_{\rm diag}={\rm diag}(M_1^2,M_2^2,M_3^2).\label{eq:massrot}
\end{equation}

\subsection{Relations among physical parameters}
\label{sec:rel}

The most general 2HDM has 11 independent real parameters. Clearly, it would be desirable to express such parameters
in terms of physical quantities, which can be measured experimentally and are, by definition, basis-invariant.
Recently~\cite{Grzadkowski:2014ada,Grzadkowski:2016szj,Grzadkowski:2018ohf} a set of 11 independent physical
parameters was proposed, described by
\begin{equation} \label{Eq:pcal}
	{\cal P}\equiv\{M_{H^\pm}^2,M_1^2,M_2^2,M_3^2,e_1,e_2,e_3,q_1,q_2,q_3,q\}.
\end{equation}
In this set, $M_{H^\pm}$ is the mass of the charged scalars, and $M_{1,2,3}$ are the masses of the three neutral scalars.  These are, in the Higgs basis, the eigenvalues of the $3\times 3$ mass matrix of
the neutral sector, diagonalized by an orthogonal matrix $R$ according to (\ref{eq:massrot}).
As for the $e_i$, they are obtained from the interactions of the neutral scalars with gauge
bosons, which arise from the doublets' kinetic terms:
\begin{equation} \label{Eq:gauge-IS}
	\lcal_k=(D_\mu \Phi_1)^\dagger(D^\mu \Phi_1) + (D_\mu \Phi_2)^\dagger(D^\mu \Phi_2).
\end{equation}
From these terms, and with the usual definitions for covariant derivatives, we identify trilinear
gauge-scalar interaction terms,
\begin{align}
	{\rm Coefficient}\left(\lcal_k,Z^\mu \left[H_j \overleftrightarrow{\partial_\mu}
	H_i\right]\right)&=\frac{g}{2v\cos\thetaW}\epsilon_{ijk}e_k,\nonumber \\
	{\rm Coefficient}\left(\lcal_k,H_i Z^\mu Z^\nu\right)&=\frac{g^2}{4\cos^2\thetaW}e_i\,g_{\mu\nu}, \nonumber \\
	{\rm Coefficient}\left(\lcal_k,H_i W^{+\mu} W^{-\nu}\right)&=\frac{g^2}{2}e_i\,g_{\mu\nu}.
\end{align}
All interactions between the $H_i$ and the electroweak gauge bosons involve the quantities
$e_i$ -- for instance, $e_1$ is related to the coupling modifier $\kappa_V$ used by the LHC experimental
collaborations by $\kappa_V = e_1/v$. In a general basis\footnote{In the Higgs basis, the expressions simplify to $e_i=vR_{i1}$.}, the $e_i$ are given by
\begin{eqnarray}
	e_i&\equiv& v_1 R_{i1}+v_2R_{i2},
\end{eqnarray}
where $R$ is the diagonalization matrix of the neutral scalars mentioned above
(see~\cite{Grzadkowski:2014ada,Grzadkowski:2016szj,Grzadkowski:2018ohf}
for details). Interestingly, the $e_i$  coefficients obey a ``sum rule"
\bea
e_1^2+e_2^2+e_3^3=v^2.
\eea
The three trilinear $H_iH^+H^-$ couplings and the quadrilinear $H^+H^+H^-H^-$ coupling complete the physical parameter set.
These couplings, respectively denoted by $q_i$ and $q$, are quite complicated in a general basis, but in the Higgs basis they simplify to
\begin{eqnarray}
	q_{i}&\equiv&{\rm Coefficient}(V,H_iH^+H^-)\label{eq:qi}
	\nonumber\\
	&=&v(R_{i1}\lambda_3+R_{i2}\Re \lambda_7-R_{i3}\Im \lambda_7),\\
	q&\equiv&{\rm Coefficient}(V,H^+H^+H^-H^-)\label{eq:q}
	\nonumber\\
	&=&\frac{1}{2}\lambda_2,
\end{eqnarray}
where again the  $R_{ij}$ are elements of the rotation matrix $R$ mentioned above.

The elements of ${\cal P}$ give therefore expressions in terms of tree-level masses and couplings,
and all physical observables of the scalar sector are expressible in terms of these 11 parameters. When symmetries are imposed
on the 2HDM the number of free parameters is reduced and relations among some of them arise. This was studied, for the six familiar symmetries of the 2HDM, in~\cite{Ferreira:2020ana}. The analysis was extended to softly
broken symmetries in~\cite{Ferreira:2022gjh}.
\subsection{The $r_0$ model}
\label{sec:r0M}

Out of all the possibilities described in table~\ref{tab:nsyms}, the $r_0$ model is the only one for which explicit CP violation
occurs. Since $m^2_{12}$ is complex, it is easy to see that its phase cannot be absorbed by a basis transformation without it rendering parameters
in the quartic part of the potential complex. Explicit CP violation can also be established using the four basis invariants whose
vanishing heralds explicit CP conservation for a given 2HDM~\cite{Gunion:2005ja}. To be more precise, we will use the equivalent formulation
of those four invariants in the bilinear formalism~\cite{Maniatis:2007vn}, given by
\begin{align}
I_1 &=\, \left(\vec{M} \times \vec{\Lambda}\right) \cdot \left(\Lambda \vec{M}\right)\nonumber \\
I_2 &=\,\left(\vec{M} \times \vec{\Lambda}\right) \cdot \left(\Lambda \vec{\Lambda}\right) \nonumber \\
I_3 &=\,\left[\vec{M} \times \left(\Lambda \vec{M}\right)\right] \cdot \left(\Lambda^2 \vec{M}\right)
\nonumber \\
I_4 &=\,\left[\vec{\Lambda} \times \left(\Lambda \vec{\Lambda}\right)\right] \cdot \left(\Lambda^2
\vec{\Lambda}\right)\,.
\label{eq:CPinv}
\end{align}
Since the $r_0$ symmetry implies $\vec{\Lambda} = \vec{0}$ the invariants $I_{1,2,4}$ are automatically zero.
This leaves $I_3$, a simple calculation shows that
\beq
I_3\,=\, -16\lambda_5\,m^2_{11}\,\mbox{Im} (m^2_{12})\,\mbox{Re} (m^2_{12})\,\left[(\lambda_1 - \lambda_3 - \lambda_4)^2 - \lambda_5^2\right],
\label{eq:I3}
\eeq
so in general we will have $I_3\neq 0$ for the $r_0$ model -- and therefore there is explicit CP violation in this model. The CP violation is not hard, but soft since the CP violating phase resides in $m_{12}^2$.

Working out the stationary-point equations for the general $r_0$ model, we find that they are solved by
\bea
m_{11}^2&=&\frac{1}{2} \lambda _1 \left(v_2^2-v_1^2\right),\nonumber\\
\Re m_{12}^2&=&\frac{1}{2} v_1 v_2 \cos \xi  \left(\lambda _1+\lambda _3+\lambda _4+\lambda _5\right),\nonumber\\
\Im m_{12}^2&=&-\frac{1}{2} v_1 v_2 \sin \xi  \left(\lambda _1+\lambda _3+\lambda _4-\lambda _5\right).
\eea
The elements of the neutral sector mass matrix become
\bea
\left({\cal M}^2\right)_{11}&=&\frac{1}{2} \left(2 \lambda _1 v_1^2+\left(\lambda _1+\lambda _3+\lambda _4+\cos 2 \xi\,  \lambda _5\right) v_2^2\right),\nonumber\\
\left({\cal M}^2\right)_{22}&=&\frac{1}{2} \left(\left(\lambda _1+\lambda _3+\lambda _4+\cos 2 \xi\,  \lambda _5\right) v_1^2+2 \lambda _1 v_2^2\right),\nonumber\\
\left({\cal M}^2\right)_{33}&=&\frac{1}{2} v^2 \left(\lambda _1+\lambda _3+\lambda _4-\cos 2 \xi\,  \lambda _5\right),\nonumber\\
\left({\cal M}^2\right)_{12}&=&-\frac{1}{2} v_1 v_2 \left(\lambda _1-\lambda _3-\lambda _4-\cos 2 \xi\,  \lambda _5\right),\nonumber\\
\left({\cal M}^2\right)_{13}&=&-\frac{1}{2}v_2v \sin 2\xi\, \lambda _5,\nonumber\\
\left({\cal M}^2\right)_{23}&=&-\frac{1}{2}v_1v \sin 2\xi\, \lambda _5.
\eea
The neutral sector rotation matrix is then given by
\bea
R=\left(
\begin{array}{ccc}
	\frac{v_2 \cos \xi }{v} & \frac{v_1 \cos \xi }{v} & -\sin \xi \\
	-\frac{v_1}{v} & \frac{v_2}{v} & 0 \\
	\frac{v_2 \sin \xi }{v} & \frac{v_1 \sin \xi }{v} & \cos \xi  \\
\end{array}
\right),
\eea
yielding masses
\begin{gather} \label{Eq:r0-masses}
M_1^2=\frac{1}{2} v^2 \left(\lambda _1+\lambda _3+\lambda _4+\lambda _5\right),\quad
M_2^2=\lambda _1 v^2, \nonumber \\
M_3^2=\frac{1}{2} v^2 \left(\lambda _1+\lambda _3+\lambda _4-\lambda _5\right),\quad
\mchsq=\frac{1}{2} \left(\lambda _1+\lambda _3\right) v^2.
\end{gather}
The $r_0$ symmetry conditions, coupled with the minimisation equations, eliminates the dependence on the
quadratic couplings. All squared masses therefore become products of quartic couplings with $v^2$. A {\em decoupling limit} is not possible in this model -- since all quartic couplings are constrained by perturbativity constraints, the values of the scalar masses cannot be too large.

With $m_\text{SM}$ the SM Higgs mass (125~GeV), and from perturbativity alone, $|\lambda_i|<4\pi$, it is easy to see from (\ref{Eq:r0-masses}) that we obtain an upper bound on the masses,
\begin{equation}
\max\{M_i\}=\sqrt{\half m_\text{SM}^2+6\pi\, v^2}\simeq 1132~\text{GeV}.
\end{equation}
Unitarity constraints\cite{Ginzburg:2005dt} on the 2HDM will however restrict the size of several combinations of quartic couplings,
so we can obtain more restrictive bounds on the scalars' masses.
A scan over parameters, imposing unitarity  and also boundedness-from-below constraints \cite{Ivanov:2006yq,Ivanov:2007de}, shows that
indeed
it is not possible to obtain scalar masses arbitrarily large, due to a combination of symmetry and unitarity conditions.
Assuming that $M_2$ is the SM-like Higgs boson, we obtain
\begin{align}
	M_{H^\pm} &\,\leq \,711 \;\mbox{GeV} \,, \nonumber \\
	M_3  &\,\leq\,712 \;\mbox{GeV} \,, \nonumber \\
	M_1 &\, \leq \,711 \;\mbox{GeV} \,,
\end{align}
and $M_1 \,+\,M_3 \leq 1400$ GeV.

Working out the three gauge couplings and the four scalar couplings contained in the physical parameter set ${\cal P}$ described in section \ref{sec:rel}, we get
\begin{gather}
e_1=\frac{2 v_1 v_2 \cos \xi }{v},\quad
e_2=\frac{v_2^2-v_1^2}{v},\quad
e_3=\frac{2 v_1 v_2 \sin \xi }{v},\nonumber\\
q_1=\frac{v_1 v_2 \cos \xi  \left(\lambda _1+\lambda _3-\lambda _4-\lambda _5\right)}{v},\quad
q_2=-\frac{\lambda _3 \left(v_1^2-v_2^2\right)}{v},\quad
q_3=\frac{v_1 v_2 \sin \xi  \left(\lambda _1+\lambda _3-\lambda _4+ \lambda _5\right)}{v},\nonumber\\
q=\frac{\lambda _1 \left(v_1^4+v_2^4\right)+2 \left(\lambda _3+\lambda _4\right) v_1^2 v_2^2++2 v_1^2 v_2^2 \cos 2 \xi \, \lambda _5}{2 v^4}.
\end{gather}
The $r_0$ conditions are easily translated into constraints among the parameters of $\pcal$. Using the techniques laid out \cite{Ogreid:2018bjq,Ferreira:2020ana}, the basis-invariant constraint $m_{11}^2+m_{22}^2=0$ translates into
\bea
\mchsq=\frac{1}{2}(e_1q_1+e_2q_2+e_3q_3)+\frac{1}{2v^2}(e_1^2M_1^2+e_2^2M_2^2+e_3^2M_3^2).\label{r0Mch}
\eea
The combined constraints $\lambda_2=\lambda_1$ and $\lambda_6+\lambda_7=0$ are also basis-invariant. These conditions are those of a CP2 invariant $V_4$, and were already translated into constraints among the parameters of $\pcal$ in \cite{Ferreira:2022gjh}, dubbed Case SOFT-CP2. Combining the constraints of Case SOFT-CP2 with (\ref{r0Mch}), we arrive at
\begin{alignat}{2}
	&\text{\bf Case $r_0$:} &\ \
	&v^2(e_1q_2-e_2q_1)+e_1e_2(M_2^2-M_1^2)=0,\quad
	v^2(e_1q_3-e_3q_1)+e_1e_3(M_3^2-M_1^2)=0,\nonumber\\
	& & &v^2(e_2q_3-e_3q_2)+e_2e_3(M_3^2-M_2^2)=0,\quad
	q=\frac{1}{2v^4}(e_1^2M_1^2+e_2^2M_2^2+e_3^2M_3^2),\nonumber\\
	& & & \mchsq=\frac{1}{2}(e_1q_1+e_2q_2+e_3q_3)+\frac{1}{2v^2}(e_1^2M_1^2+e_2^2M_2^2+e_3^2M_3^2),\nonumber
\end{alignat}
which fully describes the physical consequences of the $r_0$ symmetry when imposed upon the 2HDM potential.
Superficially, this looks like five constraints, but it is in fact only four since the first three are not independent. Thus, the most general potential invariant under $r_0$ has 11-4=7 free parameters.
It is now easy to check that the masses and couplings we worked out for this model satisfy the constraints of Case $r_0$.

\subsubsection{Soft breaking of $r_0$}
If we try to softly break $r_0$ by relaxing the condition $m_{11}^2+m_{22}^2=0$, we just go back to the softly broken CP2-model described by Case SOFT-CP2 in \cite{Ferreira:2022gjh}, except for the situation where $m_{22}^2=m_{11}^2$, and the whole potential is CP2 invariant. Such cases were described in \cite{Ferreira:2020ana}, and there only one case, namely Case CCD was found to be RG-stable.

\subsection{The 0CP1 model}
\label{sec:0CP1M}

In the 0CP1 model, the $r_0$ symmetry is imposed on the potential alongside the CP1 symmetry,
yielding a potential which, in its symmetry basis, has parameters such as are described in table~\ref{tab:nsym}, but without loss of generality we can go to a simpler basis as indicated in table~\ref{tab:nsyms} to get
\bea
V &=& m_{11}^2\left[\Phi_1^\dagger\Phi_1-\Phi_2^\dagger\Phi_2\right]
- m_{12}^2\left[\Phi_1^\dagger\Phi_2+\Phi_2^\dagger\Phi_1\right]+\half\lambda_1\left[(\Phi_1^\dagger\Phi_1)^2
+(\Phi_2^\dagger\Phi_2)^2\right]\nonumber\\
&&
+\lambda_3(\Phi_1^\dagger\Phi_1)(\Phi_2^\dagger\Phi_2)
+\lambda_4(\Phi_1^\dagger\Phi_2)(\Phi_2^\dagger\Phi_1)
+\frac{\lambda_5}{2}\left[(\Phi_1^\dagger\Phi_2)^2
+(\Phi_2^\dagger\Phi_1)^2\right]\,,
\label{eq:pot0CP1red}
\eea
where now all parameters are real. There are different ways to solve the resulting stationary point equations. There are solutions with $v_1=0$ or $v_2=0$. Such solutions imply $ m_{12}^2=0$, and are situations where the potential is $Z_2$ invariant. They will therefore not be discussed in this section. We may thus safely assume $v_1v_2\neq0$ in the following. Next, there are solutions requiring $\sin\xi=0$, thus describing a model which preserves CP, and there are also solutions where $\sin\xi\neq0$ leaving open the possibility for spontaneous CP violation.

\subsection{CP conserving 0CP1}
\label{sec:CPC0CP1}
We consider only a model with $\xi=0$ (letting $\xi=\pi$ yields similar results). Now, the stationary point equations are solved by
\bea
m_{11}^2&=&\frac{1}{2} \lambda _1 \left(v_2^2-v_1^2\right),\quad
m_{12}^2=\frac{1}{2} v_1 v_2 \left(\lambda _1+\lambda _3+\lambda _4+\lambda _5\right).
\eea
The elements of the neutral-sector mass-squared matrix are
\bea
\left({\cal M}^2\right)_{11}&=&\frac{1}{2} \left(2v_1^2\lambda_1+v_2^2 \left(\lambda_1+\lambda _3+\lambda _4+\lambda _5\right)\right),\nonumber\\
\left({\cal M}^2\right)_{22}&=&\frac{1}{2} \left(v_1^2 \left(\lambda_1+\lambda _3+\lambda _4+\lambda _5\right)+2v_2^2\lambda _1 \right),\nonumber\\
\left({\cal M}^2\right)_{33}&=&\frac{1}{2} v^2 \left(\lambda _1+\lambda _3+\lambda _4-\lambda _5\right),\nonumber\\
\left({\cal M}^2\right)_{12}&=&-\frac{1}{2} v_1 v_2 \left(\lambda _1-\lambda _3-\lambda _4-\lambda _5\right),\nonumber\\
\left({\cal M}^2\right)_{13}&=&\left({\cal M}^2\right)_{23}=0.
\eea
The rotation matrix is given by
\bea
R=\left(
\begin{array}{ccc}
	\frac{v_2}{v} & \frac{v_1}{v} & 0 \\
	-\frac{v_1}{v} & \frac{v_2}{v} & 0 \\
	0 & 0 & 1 \\
\end{array}
\right),
\eea
yielding masses
\begin{align}
M_1^2&=\frac{1}{2} v^2 \left(\lambda _1+\lambda _3+\lambda _4+\lambda _5\right),\quad
M_2^2=\lambda _1 v^2,\nonumber\\
M_3^2&=\frac{1}{2} v^2 \left(\lambda _1+\lambda _3+\lambda _4-\lambda _5\right),\quad
\mchsq=\frac{1}{2} \left(\lambda _1+\lambda _3\right) v^2.
\end{align}
The neutral-sector
mass-squared matrix is broken into two blocks -- a $2\times 2$ block, indicating the presence of two CP-even states $H_1$ and $H_2$ (this latter being the SM Higgs),
and an isolated diagonal entry indicating the mass of a pseudoscalar $H_3$. We find that
\begin{align}
	M_{H^\pm} &\,\leq \,711 \;\mbox{GeV} \,, \nonumber \\
	M_3  &\,\leq\,708 \;\mbox{GeV} \,, \nonumber \\
	M_1 &\, \leq \,961 \;\mbox{GeV} \,,
\end{align}
where we have assumed $M_2 = 125$~GeV. Once again,
we see how a decoupling limit is not achievable, as there are upper bounds on the extra scalar masses.

Working out the three gauge couplings and the four scalar couplings contained in the physical parameter set ${\cal P}$ described in section \ref{sec:rel}, we get
\begin{gather}
e_1=\frac{2 v_1 v_2}{v},\quad
e_2=\frac{v_2^2-v_1^2}{v},\quad
e_3=0,\nonumber\\
q_1=\frac{v_1 v_2 \left(\lambda _1+\lambda _3-\lambda _4-\lambda _5\right)}{v},\quad
q_2=\frac{\lambda _3 \left(v_2^2-v_1^2\right)}{v},\quad
q_3=0,\nonumber\\
q=\frac{\lambda _1 \left(v_1^4+v_2^4\right)+2 \left(\lambda _3+\lambda _4+\lambda_5\right) v_1^2 v_2^2}{2 v^4}.
\end{gather}
We see from these constraints that the model is CP conserving since $e_3=q_3=0$. This corresponds to Case C of CP conservation in \cite{Ferreira:2020ana}. Thus, combining the constraints of Case $r_0$ with the constraints of Case C, we arrive at
\begin{alignat}{2}
	&\text{\bf Case 0CP1-C:} &\ \
	&  e_k=q_k=0,\quad v^2(e_iq_j-e_jq_i)+e_ie_j(M_j^2-M_i^2)=0, \nonumber\\
	& & & 
	q=\frac{1}{2v^4}(e_1^2M_1^2+e_2^2M_2^2),\quad
	\mchsq=\frac{1}{2}(e_1q_1+e_2q_2)+\frac{1}{2v^2}(e_1^2M_1^2+e_2^2M_2^2)\nonumber
\end{alignat}
which fully describes the physical consequences of the CP conserving 0CP1 model. There are five constraints, implying that this model has 11-5=6 free parameters.
It is now easy to check that the masses and couplings we worked out for this model satisfy the constraints of Case 0CP1-C for $k=3$.

\subsubsection{Spontaneous CP violation in a 0CP1 model}

For the regular CP1-conserving potential, we know that, for certain
regions of parameter space, spontaneous CP violation (SCPV) can occur, so let us investigate whether the same can happen
for the 0CP1 model. Solving the stationary point equations assuming $\sin\xi\neq0$ yields
\bea
m_{11}^2&=&\frac{1}{2} \lambda _1 \left(v_2^2-v_1^2\right),\\
m_{12}^2&=&\left(\lambda _1+\lambda _3+\lambda _4\right) v_1 v_2 \cos \xi ,\\
\lambda _5&=&\lambda _1+\lambda _3+\lambda _4.\label{unstable}
\eea
The last of these equations is a condition among quartic couplings only, not
enforced by the model's symmetries, and which therefore would be a tree-level fine-tuning,
unstable 	under radiative corrections. Since we are assuming that $v_1v_2\neq0$, as otherwise
that would imply a $Z_2$ invariant vacuum,  the only way to avoid RG-instability is to assume $\sin\xi=0$,
as we did in section \ref{sec:CPC0CP1}. Notice, also, that in the  situation encountered, the minimization
conditions do not allow for a full determination of the parameters $v_1$, $v_2$ and $\xi$ in terms of potential
parameters. This is presumably a situation where the tree-level minimisation conditions are not sufficient to
determine whether SCPV can occur, and one would need to perform a one-loop analysis to settle the issue.
We will meet this issue again, for the 0$Z_2$ and 0U(1) models. The only vacua in the model for which we can rely on
the tree-level solutions are therefore those which preserve CP, i.e., with $\sin\xi=0$.
We will therefore not investigate this model further in the present work, since a full one-loop analysis
is needed to settle the issue of SCPV.

\subsubsection{Soft breaking of CP1}
Let us consider the possibility of keeping the $r_0$ symmetry intact, but softly breaking CP1. That would imply that we allow for complex $m_{12}^2$. From table \ref{tab:nsyms} we see that this simply takes us back to the general $r_0$ model and yields nothing new.
\subsection{The 0$Z_2$ model}
\label{sec:0Z2M}

As seen from table~\ref{tab:nsyms}, the 0$Z_2$ model is characterized (in the reduced basis) by, on top of the relations
between parameters from the 0CP1 model, also having $m^2_{12} = 0$. The potential then reads
\bea
V &=& m_{11}^2\left[\Phi_1^\dagger\Phi_1-\Phi_2^\dagger\Phi_2\right]
+\half\lambda_1\left[(\Phi_1^\dagger\Phi_1)^2
+(\Phi_2^\dagger\Phi_2)^2\right]
+\lambda_3(\Phi_1^\dagger\Phi_1)(\Phi_2^\dagger\Phi_2)\nonumber\\[8pt]
&&\quad
+\lambda_4(\Phi_1^\dagger\Phi_2)(\Phi_2^\dagger\Phi_1)
+\frac{\lambda_5}{2}\left[(\Phi_1^\dagger\Phi_2)^2
+(\Phi_2^\dagger\Phi_1)^2\right]\,,
\label{eq:pot0Z2red}
\eea
There are different ways to solve the stationary point equations. There are solutions with $v_1=0$ or $v_2=0$. Such solutions imply that the model is $Z_2$ invariant (inert).
Next, there are solutions requiring $\sin2\xi=0$. They represent models where $Z_2$ may be spontaneously broken.
There are also solutions where $v_1v_2\neq0$ and $\sin 2\xi\neq0$. Such solutions will imply $\lambda_5=0$, and this yields a U(1) invariant potential. Such solutions will therefore not be discussed in this section.

\subsubsection{$Z_2$ conserving vacuum in 0$Z_2$}

We consider only a model with $v_2=0$ (letting $v_1=0$ yields similar results). We may then without loss of generality rotate to a basis where $\xi=0$. Now, the stationary point equations are solved by
\bea
m_{11}^2=-\frac{1}{2} \lambda _1 v^2.
\eea
The neutral sector mass matrix is diagonal and without mass degeneracy, so
the rotation matrix is simply $R=I_3$,
and masses are given by
\begin{align}
M_1^2&=\lambda _1 v^2,\quad
M_2^2=\frac{1}{2} v^2 \left(\lambda _1+\lambda _3+\lambda _4+\lambda _5\right),\nonumber\\
M_3^2&=\frac{1}{2} v^2 \left(\lambda _1+\lambda _3+\lambda _4-\lambda _5\right),\quad
\mchsq=\frac{1}{2} \left(\lambda _1+\lambda _3\right) v^2.
\end{align}
The vacuum with $v_2 = 0$ we just came across is clearly the realization, within the 0$Z_2$ model, of
the Inert 2HDM~\cite{Ma:1978,Barbieri:2006,Cao:2007,LopezHonorez:2006}. In such a model, the observed Higgs
boson ($h\equiv H_1$) stems from the real, neutral
component of $\Phi_1$ and has (tree-level) interactions with gauge bosons (and fermions) identical to
those of the SM, whereas the extra scalars arising from $\Phi_2$ -- which would be $H\equiv H_2$, $A\equiv H_3$ and $H^\pm$ --
have no triple vertex interactions with gauge bosons (and fermions). The lightest of those states is therefore
stable and the resulting IDM has been studied extensively as a possible model to provide dark
matter.\footnote{Obviously, we could also have a vacuum with $v_1 = 0$,
	but in what concerns the scalar sector that solution is equivalent, via a basis change, to the $v_2 = 0$
	solution. When fermions are taken into account they are not, however, equivalent.}

A quick scan demanding unitarity and boundedness from below for the quartic couplings yields an upper bound
of roughly 710 GeV for all extra scalar (non-SM Higgs states, therefore) masses. This is in stark contrast with
the usual IDM, for which there is no upper bound for the inert scalar masses, since the $m^2_{22}$ parameter
is free in that model. But not here -- the $r_0$ symmetry forces  $m^2_{22} = -m^2_{11} = M_1^2/2$, and
thus upper bounds on the $H$, $A$ and $H^\pm$ arise.

Working out the three gauge couplings and the four scalar couplings contained in the physical parameter set ${\cal P}$ described in section \ref{sec:rel}, we get
\bea
e_1=v,\quad
e_2=e_3=0,\quad
q_1=\lambda _3 v,\quad
q_2=q_3=0,\quad
q=\frac{\lambda _1}{2}.
\eea
We see that the model is $Z_2$ invariant since $e_2=e_3=q_2=q_3=0$. This corresponds to Case CC of $Z_2$ conservation in \cite{Ferreira:2020ana}. Thus, combining the constraints of Case $r_0$ with the constraints of Case CC, we arrive at
\begin{alignat}{2}
	&\text{\bf Case 0$Z_2$-CC:} &\ \
	&e_j=q_j=e_k=q_k=0,\quad q=\frac{M_i^2}{2v^2},\quad
	 \mchsq=\frac{e_iq_i}{2}+\frac{M_i^2}{2},\nonumber
\end{alignat}
which fully describes the physical consequences of the $Z_2$ invariant 0$Z_2$ model. There are six constraints, implying that this model has has 11-6=5 free parameters.
It is now easy to check that the masses and couplings we worked out for this model satisfy the constraints of Case 0$Z_2$-CC for $i=1$, $j=2$ and $k=3$.

\subsubsection{Spontaneous $Z_2$ violation in a 0$Z_2$ model}

We solve the stationary point equations for $\sin2\xi=0$, assuming $v_1v_2\neq0$. We restrict ourselves to $\xi=0$ ($\xi=\pm\pi/2,\pi$ yields similar results). We get
\bea
m_{11}^2=\frac{1}{2} \lambda _1 \left(v_2^2-v_1^2\right),\quad
\lambda _5=-\lambda _1-\lambda _3-\lambda _4.
\eea
Again, the last of these equations is a condition among quartic couplings only, not
enforced by the model's symmetries, and which therefore would be a tree-level fine-tuning, unstable
under radiative corrections. See the discussion following (\ref{unstable}). The tree-level minimization conditions
are not letting us determine both vevs in terms of the potential parameters, only $v_1^2-v_2^2$ could be found.
Note also that the condition $\lambda_5=-\lambda_1 - \lambda_3 - \lambda_4$ is not preserved by the RGE, i.e. the corresponding $\beta$-function is not
vanishing. So once again, a one-loop minimization is necessary to investigate the possibility of $v_2\neq 0$ -- we can expect that higher orders of perturbation expansion are necessary to generate a non-zero true vev somewhere along the tree-level one in accordance with the Georgi-Pais theorem~\cite{Georgi:1974au}.
Since again a full one-loop analysis is necessary, we will not investigate this model further here.

\subsubsection{Soft breaking of $Z_2$}

Let us also consider the possibility of keeping the $r_0$ symmetry intact, but softly breaking $Z_2$. That would imply that we allow for a nonzero $m_{12}^2$. From table \ref{tab:nsyms} we see that this simply takes us back to the 0CP1 model in the case of a real $m_{12}^2$ and to the general $r_0$ model in the case of a complex $m_{12}^2$, hence this yields nothing new.

\subsection{The 0U(1) model}
\label{sec:0U1M}

As can be appreciated from table~\ref{tab:nsyms}, the 0U(1) has in the reduced basis all the parameter constraints of the 0$Z_2$ one, plus the
condition $\lambda_5 = 0$. The potential reads
\bea
V &=& m_{11}^2\left[\Phi_1^\dagger\Phi_1-\Phi_2^\dagger\Phi_2\right]
+\half\lambda_1\left[(\Phi_1^\dagger\Phi_1)^2
+(\Phi_2^\dagger\Phi_2)^2\right]\nonumber\\
&&
+\lambda_3(\Phi_1^\dagger\Phi_1)(\Phi_2^\dagger\Phi_2)
+\lambda_4(\Phi_1^\dagger\Phi_2)(\Phi_2^\dagger\Phi_1),
\label{eq:potU1}
\eea
and we may without loss of generality rotate into a basis where $\xi=0$. There are different ways to solve the stationary point equations. There are solutions with $v_1=0$ or $v_2=0$. Such solutions imply that the whole model is U(1) invariant.
There are also solutions where $v_1v_2\neq0$. In such models, U(1) is spontaneously broken.

\subsubsection{U(1) invariant vacuum in 0U(1)}

We consider only a model with $v_2=0$ (letting $v_1=0$ yields similar results). Now, the stationary point equations are solved by
\bea
m_{11}^2=-\frac{1}{2} \lambda _1 v^2.
\eea
The neutral sector mass matrix is diagonal with masses given by\footnote{
Since there is mass degeneracy one can imagine mixing the two mass degenerate states $H_2$ and $H_3$ using a rotation matrix 	
$
R=\left(
\begin{array}{ccc}
	1 & 0 & 0 \\
	0 & \cos\alpha & \sin\alpha \\
	0 & -\sin\alpha & \cos\alpha \\
\end{array}
\right),$
where $\alpha$ is completely arbitrary. Note that none of the masses or couplings depend on $\alpha$, so simply putting $\alpha=0$ yields the exact same result.}
\bea
M_1^2=\lambda _1 v^2,\quad
M_2^2=M_3^2=\frac{1}{2} \left(\lambda _1+\lambda _3+\lambda _4\right) v^2,\quad
\mchsq=\frac{1}{2} \left(\lambda _1+\lambda _3\right) v^2.
\eea
We conclude that if the U(1) symmetry is preserved by the vacuum (when only one of the doublets acquires a vev) we are left
with a version of the IDM, where the two neutral inert scalars are degenerate in mass.
Working out the three gauge couplings and the four scalar couplings contained in the physical parameter set ${\cal P}$ described in section \ref{sec:rel}, we get
\bea
e_1=v,\quad
e_2=e_3=0,\quad
q_1=\lambda _3 v,\quad
q_2=q_3=0,\quad
q=\frac{\lambda _1}{2}.
\eea
We see that the model is U(1) conserving since all the constraints defining Case BCC of U(1) conservation in \cite{Ferreira:2020ana} are satisfied. Thus, combining the constraints of Case $r_0$ with the constraints of Case BCC, we arrive at
\begin{alignat}{2}
	&\text{\bf Case 0U(1)-BCC:} &\ \
	&M_j=M_k,\quad e_j=q_j=e_k=q_k=0,\quad q=\frac{M_i^2}{2v^2},\nonumber\\
	& & & \mchsq=\frac{e_iq_i}{2}+\frac{M_i^2}{2},\nonumber
\end{alignat}
which fully describes the physical consequences of the U(1) conserving 0U(1) model. There are seven constraints, implying that this model has has 11-7=4 free parameters.
It is now easy to check that the masses and couplings we worked out for this model satisfy the constraints of Case 0U(1)-BCC for $i=1$, $j=2$ and $k=3$.

\subsubsection{Spontaneous U(1) violation in a 0U(1) model}
We solve the stationary point equations for $v_1v_2\neq0$,
\bea
m_{11}^2=\frac{1}{2} \lambda _1 \left(v_2^2-v_1^2\right),\quad
\lambda _4=-\lambda _1-\lambda _3.
\eea
Again, $\lambda_4=-\lambda_1 - \lambda_3$ is an RGE unstable condition. Furthermore, this vacuum would leave
undetermined the values of the vevs $v_1$ and $v_2$. As in the previous RGE-unstable cases encountered, a one-loop calculation would be necessary to investigate the possibility of spontaneous 0U(1) breaking.
We will not pursue this model further in the present work, a full one-loop analysis is needed to settle the issue of spontaneous breaking of U(1).

\subsubsection{Soft breaking of 0U(1)}

Let us consider the possibility of keeping the $r_0$ symmetry intact, but softly break U(1). That would imply that we allow for a nonzero $m_{12}^2$.
The potential for an $r_0$ invariant potential with a softly broken U(1) is
\bea
V &=& m_{11}^2\left[\Phi_1^\dagger\Phi_1-\Phi_2^\dagger\Phi_2\right]
-\left[m_{12}^2\Phi_1^\dagger\Phi_2+{\rm h.c.}\right]
+\half\lambda_1\left[(\Phi_1^\dagger\Phi_1)^2
+(\Phi_2^\dagger\Phi_2)^2\right]\nonumber\\
&&
+\lambda_3(\Phi_1^\dagger\Phi_1)(\Phi_2^\dagger\Phi_2)
+\lambda_4(\Phi_1^\dagger\Phi_2)(\Phi_2^\dagger\Phi_1).
\label{eq:pot0U1}
\eea
Without loss of generality we may rotate into a basis in which $m_{12}^2$ is real to get
\bea
V &=& m_{11}^2\left[\Phi_1^\dagger\Phi_1-\Phi_2^\dagger\Phi_2\right]
-m_{12}^2\left[\Phi_1^\dagger\Phi_2+\Phi_2^\dagger\Phi_1\right]
+\half\lambda_1\left[(\Phi_1^\dagger\Phi_1)^2
+(\Phi_2^\dagger\Phi_2)^2\right]\nonumber\\
&&
+\lambda_3(\Phi_1^\dagger\Phi_1)(\Phi_2^\dagger\Phi_2)
+\lambda_4(\Phi_1^\dagger\Phi_2)(\Phi_2^\dagger\Phi_1).
\label{eq:pot0U1red}
\eea
The only viable vacua require $\sin\xi=0$. We choose to analyze the situation where $\xi=0$ (if $\xi=\pi$ we get similar results).
The stationary point equations are then solved by
\bea
m_{11}^2=\frac{1}{2} \lambda _1 \left(v_2^2-v_1^2\right),\quad
\Re m_{12}^2=\frac{1}{2} \left(\lambda _1+\lambda _3+\lambda _4\right) v_1 v_2.
\eea
The neutral sector mass matrix is given by
\bea
\frac{1}{2}\left(
\begin{array}{ccc}
	2\lambda _1 v_1^2+\left(\lambda_1+\lambda _3+\lambda _4\right) v_2^2 &  \left(-\lambda _1+\lambda _3+\lambda _4\right) v_1 v_2 & 0 \\
	 \left(-\lambda _1+\lambda _3+\lambda _4\right) v_1 v_2 &  \left(\lambda_1+\lambda _3+\lambda _4\right) v_1^2+2\lambda _1  v_2^2 & 0 \\
	0 & 0 &  \left(\lambda _1+\lambda _3+\lambda _4\right) v^2 \\
\end{array}
\right),
\eea
and since there is mass degeneracy,
the most general rotation matrix is given by
\bea
R=\left(
\begin{array}{ccc}
	\frac{v_2 \cos \alpha }{v} & \frac{v_1 \cos \alpha }{v} & \sin \alpha  \\
	-\frac{v_1}{v} & \frac{v_2}{v} & 0 \\
	-\frac{v_2 \sin \alpha }{v} & -\frac{v_1 \sin \alpha }{v} & \cos \alpha  \\
\end{array}
\right),
\eea
where $\alpha$ is arbitrary (and simply mixes the two mass degenerate fields $H_1$ and $H_3$).
Masses are given as
\bea
M_1^2=M_3^2=\frac{1}{2} \left(\lambda _1+\lambda _3+\lambda _4\right) v^2,\quad
M_2^2=\lambda _1 v^2,\quad
\mchsq=\frac{1}{2}(\lambda_1+\lambda_3)v^2,
\eea
and the couplings are
\begin{gather}
e_1=\frac{2 v_1 v_2 \cos \alpha }{v},\quad
e_2=\frac{v_2^2-v_1^2}{v},\quad
e_3=-\frac{2 v_1 v_2 \sin \alpha }{v},\nonumber\\
q_1=\frac{\left(\lambda _1+\lambda _3-\lambda _4\right) v_1 v_2 \cos \alpha }{v},\quad
q_2=\frac{\lambda _3 \left(v_2^2-v_1^2\right)}{v},\quad
q_3=-\frac{\left(\lambda _1+\lambda _3-\lambda _4\right) v_1 v_2 \sin \alpha }{v},\nonumber\\
q=\frac{\lambda _1 \left(v_1^4+v_2^4\right)+2 \left(\lambda _3+\lambda _4\right) v_1^2 v_2^2}{2 v^4}.
\end{gather}
Now it is easy to check that the physical constraints are satisfied for this model with $i=1,j=3,k=2$:
\begin{alignat}{2}
	&\text{\bf Case SOFT-0U1-B:} &\ \
	&M_i=M_j,\quad
	e_iq_j-e_jq_i=0,\nonumber\\
	& & &v^2(e_iq_k-e_kq_i)+e_ie_k(M_k^2-M_i^2)=0,\nonumber\\
	& & &v^2(e_jq_k-e_kq_j)+e_je_k(M_k^2-M_i^2)=0,\nonumber\\
	& & & q=\frac{1}{2v^4}([e_i^2+e_j^2]M_i^2+e_k^2M_k^2),\nonumber\\
	& & & \mchsq=\frac{1}{2}([e_iq_i+e_jq_j]+e_kq_k)+\frac{1}{2v^2}([e_i^2+e_j^2]M_i^2+e_k^2M_k^2),\nonumber
\end{alignat}
and the presence of Case B \cite{Ferreira:2020ana} of CP conservation tells us that CP violation is not possible.

Notice that, interestingly,  one obtains a degeneracy (at tree-level) between two of the neutral states, both in the case of the inert 0U(1)
model and the softly broken version of 0U(1). Indeed, both models have analogous expressions for the masses, but there is a crucial
distinction between them: in the inert 0U(1) model, only one of those scalars (denoted $H_1$) will have tree-level couplings to $W$ and $Z$ pairs,
whereas the others ($H_2$ and $H_3$) are
indeed inert states -- thus, neither $H_2$ nor $H_3$ couple to electroweak gauge bosons at tree level. In the softly broken 0U(1) model, for
which both doublets have vevs, the CP-even mass matrix is not diagonal in the
symmetry basis, indicating that mixing occurs between the CP-even parts of the two doublets.
Also, we see that some couplings depend on the arbitrary angle $\alpha$. In \cite{Ferreira:2020ana}, we argued that in such situations, what we can observe in experiments cannot depend on the unphysical $\alpha$, only combinations independent of $\alpha$ may
 appear\footnote{For instance combinations $e_1^2+e_3^2$, $q_1^2+q_3^2$, $e_1q_1+e_3q_3$  are independent of $\alpha$.}.
 Nevertheless, we can pick a particular value of $\alpha$ and perform our analysis and calculations of observables with the
 chosen value of $\alpha$. Picking $\alpha=0$ leads to $e_3=q_3=0$ and identifies $H_3$ as a pseudoscalar that does not couple
 to CP-even pairs of gauge bosons ($ZZ$, $W^+W^+$) or charged scalars ($H^+H^-$). Therefore, though degenerate at tree-level,
 $H_1$ and $H_3$ have different interactions, which indicates that their mass degeneracy will be lifted by radiative corrections.
This argument cannot hold for one value of $\alpha$ only, but holds irrespective of the value of $\alpha$ one chooses.

There is also a sub-case of SOFT-0U1-B that we get if we put $m_{11}^2=0$. The only viable vacuum then is whenever $\sin\xi=0$ and $v_1=v_2=v/\sqrt{2}$. The analysis is identical to the steps above, leading to
\begin{gather}
	e_1=v \cos \alpha,\quad
	e_2=0,\quad
	e_3=-v \sin \alpha ,\nonumber\\
	q_1=\half\left(\lambda _1+\lambda _3-\lambda _4\right) v \cos \alpha ,\quad
	q_2=0,\quad
	q_3=-\half\left(\lambda _1+\lambda _3-\lambda _4\right) v \sin \alpha ,\nonumber\\
	q=\frac{1}{4}\left(\lambda _1+\lambda _3+\lambda _4\right).
\end{gather}
Now it is easy to check that the physical constraints are satisfied for this model with $i=1,j=3,k=2$:
\begin{alignat}{2}
	&\text{\bf Case SOFT-0U1-BC:} &\ \
	&M_i=M_j,\quad
	e_iq_j-e_jq_i=0,\quad e_k=q_k=0\nonumber\\
	& & & q=\frac{M_i^2}{2v^2},\quad
	\mchsq=\frac{1}{2}(e_iq_i+e_jq_j)+\frac{M_i^2}{2}.\nonumber
\end{alignat}
Note that this is the same model that one gets if one in the softly broken 0CP3 model of (\ref{eq:0CP3softred}) considers the sub-case where $m_{12}^2$ is real. As is shown later, the softly broken 0U(1) models and the softly broken 0CP3 models are simply related via a change of basis.
\subsection{The 0CP2 model}
In the 0CP2 model there are no quadratic terms, therefore no spontaneous electroweak breaking may occur for those
cases (at tree-level, at least). Adding soft terms that break CP2 while keeping the $r_0$ symmetry intact simply takes us back to the $r_0$, 0CP1 or 0$Z_2$ model (depending on whether $m_{12}^2$ is complex, real  or vanishing) as can be seen from table \ref{tab:nsyms}. Hence, there are no new realistic models to be found by studying 0CP2 models.

\subsection{The 0CP3 model (softly broken)}

In the 0CP3 model there are no quadratic terms as well, therefore no spontaneous electroweak breaking may occur for those
cases (at tree-level, at least). Adding soft terms that break CP3 while keeping the $r_0$ symmetry intact
yields the following potential
\bea
V &=& m_{11}^2\left[\Phi_1^\dagger\Phi_1-\Phi_2^\dagger\Phi_2\right]
-\left[m_{12}^2\Phi_1^\dagger\Phi_2+{\rm h.c.}\right]
+\half\lambda_1\left[(\Phi_1^\dagger\Phi_1)^2
+(\Phi_2^\dagger\Phi_2)^2\right]\nonumber\\
&&
+\lambda_3(\Phi_1^\dagger\Phi_1)(\Phi_2^\dagger\Phi_2)
+\lambda_4(\Phi_1^\dagger\Phi_2)(\Phi_2^\dagger\Phi_1)
+ \frac{\lambda_1-\lambda_3-\lambda_4}{2}\left[(\Phi_1^\dagger\Phi_2)^2 + (\Phi_2^\dagger\Phi_1)^2\right].
\label{eq:0CP3soft}
\eea
Without loss of generality we can employ a change of basis with an orthogonal rotation among the two doublets, with a choice of either making $m_{11}^2=0$ or making $m_{12}^2$ purely imaginary to further simplify the potential ($m_{12}^2$ cannot be made real using an orthogonal change of basis). We choose to simplify the potential further by making $m_{11}^2=0$  to get
\bea
V &=&
-\left[m_{12}^2\Phi_1^\dagger\Phi_2+{\rm h.c.}\right]
+\half\lambda_1\left[(\Phi_1^\dagger\Phi_1)^2
+(\Phi_2^\dagger\Phi_2)^2\right]\nonumber\\
&&
+\lambda_3(\Phi_1^\dagger\Phi_1)(\Phi_2^\dagger\Phi_2)
+\lambda_4(\Phi_1^\dagger\Phi_2)(\Phi_2^\dagger\Phi_1)
+ \frac{\lambda_1-\lambda_3-\lambda_4}{2}\left[(\Phi_1^\dagger\Phi_2)^2 + (\Phi_2^\dagger\Phi_1)^2\right].
\label{eq:0CP3softred}
\eea
This would seem a completely new possibility, but indeed it is not -- it is in fact the same
potential of the softly-broken 0U(1) model, eq.~\eqref{eq:pot0U1red}, but expressed in a different basis. To
see this, start from that equation and use the expressions for basis changes shown in section~\ref{sec:basis}, for
the following basis transformation:
\beq
\begin{pmatrix} \Phi^\prime_1 \\ \Phi^\prime_2 \end{pmatrix}\,=\,
\frac{1}{\sqrt{2}}\,
\begin{pmatrix} 1 & -i \\ -i & 1 \end{pmatrix}
\,\begin{pmatrix} \Phi_1 \\ \Phi_2 \end{pmatrix}\,.
\eeq
In the new basis, the potential will have the exact form of eq.~\eqref{eq:0CP3softred}. This case therefore
yields nothing new.

\subsection{The 0SO(3) model (softly broken)}

In the 0SO(3) model there are again no quadratic terms, therefore no spontaneous electroweak breaking may occur for those
cases (at tree-level, at least). Adding soft terms that break SO(3) while keeping the $r_0$ symmetry intact
yield the following potential
\bea
V &=& m_{11}^2\left[\Phi_1^\dagger\Phi_1-\Phi_2^\dagger\Phi_2\right]
-\left[m_{12}^2\Phi_1^\dagger\Phi_2+{\rm h.c.}\right]
+\half\lambda_1\left[(\Phi_1^\dagger\Phi_1)^2
+(\Phi_2^\dagger\Phi_2)^2\right]\nonumber\\
&&
+\lambda_3(\Phi_1^\dagger\Phi_1)(\Phi_2^\dagger\Phi_2)
+(\lambda_1-\lambda_3)(\Phi_1^\dagger\Phi_2)(\Phi_2^\dagger\Phi_1).
\label{eq:0SO3soft}
\eea
Since the quartic part of the SO(3) potential is insensitive to basis changes, we may without loss of generality rotate into a basis where $m_{12}^2=0$ to get
\bea
V &=& m_{11}^2\left[\Phi_1^\dagger\Phi_1-\Phi_2^\dagger\Phi_2\right]
+\half\lambda_1\left[(\Phi_1^\dagger\Phi_1)^2
+(\Phi_2^\dagger\Phi_2)^2\right]\nonumber\\
&&
+\lambda_3(\Phi_1^\dagger\Phi_1)(\Phi_2^\dagger\Phi_2)
+(\lambda_1-\lambda_3)(\Phi_1^\dagger\Phi_2)(\Phi_2^\dagger\Phi_1).
\label{eq:0SO3softred}
\eea
We may also without loss of generality assume $\xi=0$.
The only viable solution\footnote{Another solution with $m_{11}^2 =0$, $\lambda_1=0$ also exist, but then we are back to the situation where we have no quadratic terms, so electroweak symmetry breaking does not occur.} of the stationary point equations is found for $v_2=0$ ($v_1=0$ yields similar results),
\bea
m_{11}^2= -\frac{1}{2} \lambda _1 v^2,\quad
v_2=0.
\eea
The mass matrix is diagonal with full mass degeneracy.
\bea
M_1^2=M_2^2=M_3^2=\lambda _1 v^2,\quad
\mchsq=\frac{1}{2}(\lambda_1+\lambda_3)v^2.
\eea
The most general rotation matrix is therefore given by
\bea
R=\begin{pmatrix}
	c_1\,c_2 & s_1\,c_2 & s_2 \\
	- (c_1\,s_2\,s_3 + s_1\,c_3)
	& c_1\,c_3 - s_1\,s_2\,s_3 & c_2\,s_3 \\
	- c_1\,s_2\,c_3 + s_1\,s_3
	& - (c_1\,s_3 + s_1\,s_2\,c_3) & c_2\,c_3
\end{pmatrix},\label{fullrotationmatrix}
\eea
where $c_i=\cos\alpha_i$, $s_i=\sin\alpha_i$ and all $\alpha_i$ are completely arbitrary due to the full mass degeneracy.

The couplings are
\begin{gather}
e_1=v c_1\,c_2 ,\quad
e_2=-v (c_1\,s_2\,s_3 + s_1\,c_3) ,\quad
e_3=v(- c_1\,s_2\,c_3 + s_1\,s_3) ,\nonumber\\
q_1=v\lambda _3c_1\,c_2 ,\quad
q_2=-v \lambda _3   (c_1\,s_2\,s_3 + s_1\,c_3) ,\quad
q_3=v\lambda _3 (- c_1\,s_2\,c_3 + s_1\,s_3) ,\quad
q=\frac{\lambda _1}{2}.
\end{gather}
This model was discussed in \cite{Ferreira:2022gjh} where we dubbed it Case SOFT-SO3-ABBB. Here, we add  a zero to the name since it is invariant under $r_0$. In terms of masses and couplings, this model is then described by
\begin{alignat}{2}
&\text{\bf Case SOFT-0SO3-ABBB:} &\ \
&M_1=M_2=M_3,\quad
e_1q_2-e_2q_1=0,\quad e_1q_3-e_3q_1=0,\quad e_2q_3-e_3q_2=0,\nonumber\\
& & &2M_{H^\pm}^2=M_1^2+e_1q_1+e_2q_2+e_3q_3,\quad2v^2 q=M_1^2.\nonumber
\end{alignat}
We may also here pick specific values of the arbitrary rotation angles. Picking all $\alpha_i=0$, yields $e_2=e_3=q_2=q_3=0$, thereby identifying $H_2$ and $H_3$ as the inert scalars that do not couple to CP-even pairs of gauge bosons ($ZZ$, $W^+W^+$) or charged scalars ($H^+H^-$). Since then $H_1$ couples differently to the gauge bosons than the inert fields $H_2$ and $H_3$ do, we expect the full mass degeneracy to be lifted at one-loop level. A partial mass degeneracy between the two inert fields $H_2$ and $H_3$ may very well be preserved at one loop level.


\section{The fermion sector}
\label{sec:ferm}

We have established in previous sections that the conditions described by eqs.~\eqref{eq:nsc} are RG invariant to all orders. Our
demonstrations, however, involved only the scalar and gauge sectors. That by itself is interesting, as we may consider conceptual models
without fermions, but as we will now show, the conditions behind the $r_0$ symmetry (as well as several other of the new symmetries studied
above) can be satisfied to at least two-loop order, even if one includes the Yukawa sector. This is more than can be said, for instance,
for the ``custodial symmetry", which not only is broken by the $U(1)_Y$ gauge group, but also by the different masses for up and down
quarks. In this section we do not wish to exhaust all possibilities, but simply show that it is possible to find Yukawa textures
which are invariant, up to two-loop order, under some of the symmetries discussed earlier.

Concerning the new symmetries proposed in the current work, whose effects on the parameters of the potential are summarised in
table~\ref{tab:nsym}, we observe that they all have a scalar quartic sector with couplings which obey, {\em at least}, the CP2
symmetry relations. For models with symmetries such as $r_0$, 0CP1 and 0CP2, indeed, the quartic sector obeys exactly the same relations
as the CP2 case. This means that, if we can find Yukawa matrices with textures which comply with the CP2 symmetry, we automatically will
have ensured that:
\begin{itemize}
\item Those textures will be preserved under radiative corrections, since they are the result of a symmetry (CP2) which extends to
all dimensionless couplings of the model.
\item The relations between quartic scalar couplings in (at least) models $r_0$, 0CP1 and 0CP2 (softly broken or not) will
be preserved to all orders in perturbation theory.
\item The theory will be renormalizable regardless of the quadratic parameters of the scalar potential, but it may be possible that
the $r_0$ relation $m^2_{22} = - m^2_{11}$ is RG-preserved even when considering Yukawa interactions.
\end{itemize}
In other words, in what concerns the dimensionless couplings of the model (scalar quartic, gauge or Yukawa), choosing CP2
Yukawa textures is consistent from the renormalization point of view: CP2 Yukawas will not spoil the 0CP2 scalar quartic
relations because they are identical to the CP2 ones, and vice-versa. It remains to be seen whether CP2 Yukawas respect the full
0CP2 symmetry-imposed relations, {\em i.e.,} the relation $m^2_{22} = - m^2_{11}$. We will show that this is what happens,
at least up to two-loop order.

The same arguments are valid if we consider the 0CP3 model (softly broken or not) -- since that model has quartic coupling relations which are identical
to the CP3 case, if one considers a CP3-symmetric Yukawa sector
all relations between dimensionless couplings are preserved under renormalization. Again, it remains also a possibility that the $m^2_{22} = - m^2_{11}$ relation
is itself found to be preserved under radiative corrections. It is this aspect which we will now investigate, since this relation
between quadratic parameters is what distinguishes the new symmetries we are proposing from those already known.

Let us recall that the most generic 2HDM Yukawa sector may be written as\footnote{We will neglect neutrinos in this
study; pure Dirac mass terms for neutrinos could be trivially added to this lagrangian, of course.}
\beq
- {\cal L}_Y \,=\, \bar{q}_L (\Gamma_1 \Phi_1 + \Gamma_2 \Phi_2) n_R \,+\, \bar{q}_L (\Delta_1 \tilde{\Phi}_1 + \Delta_2 \tilde{\Phi}_2) p_R
\, +\, \bar{l}_L (\Pi_1 \Phi_1 + \Pi_2 \Phi_2) l_R \, +\, \textrm{H.c.}
\eeq
In this equation, $\tilde{\Phi}_i = i\sigma_2\Phi_i^*$ are the doublets' charge conjugates; $q_L$ and $l_L$ are 3-vectors in flavour space
containing the quark and lepton left doublets; likewise, $n_R$, $p_R$ and $l_R$ are 3-vectors in flavour space, containing, respectively, the
righthanded down, up and charged lepton fields. The $\Gamma_i$, $\Delta_i$ and $\Pi_i$ are $3\times 3$ complex matrices containing Yukawa couplings.
The fermionic fields in this equation do not correspond to the quark and lepton mass states. The physical fields (corresponding to
quark and lepton mass eigenstates) are related to these via unitary transformations in flavour space which involve $3\times 3$ U(3) matrices in
flavour space. For the quarks, for instance, we would have
\beq
p_L = U_{uL} \,u_L\;\;,\;\;  p_R = U_{uR}\, u_R \;\;,\;\; \,,  n_L = U_{dL} \,d_L\;\;,\;\;n_R = U_{dR}\, d_R  
\label{eq:qbc}
\eeq
so that the down and up quark mass matrices, given by
\beq
M_d\,=\,\frac{1}{\sqrt{2}}\,\left(\Gamma_1 v_1 \,+\, \Gamma_2 v_2\right)\;\; , \;\;
M_u\,=\,\frac{1}{\sqrt{2}}\,\left(\Delta_1 v_1^* \,+\, \Delta_2 v_2^*\right)
\eeq
are bi-diagonalised so that one obtains the physical quark masses,
\beq
\textrm{diag}(m_d\,,\,m_s\,,\,m_b)\,=\, U_{dL}^{\dagger}\,M_d\,U_{dR}\;\;\;,\;\;\;
\textrm{diag}(m_u\,,\,m_c\,,\,m_t)\,=\, U_{uL}^{\dagger}\,M_u\,U_{uR}\,.
\eeq
Similar relations hold for the leptons as well. The transformations~\eqref{eq:qbc} mean that there is additional
basis freedom in the Yukawa sector of the 2HDM, by redefining the fermion fields alongside the scalar ones. Concerning
the CP symmetries -- CP1, CP2 and CP3 -- of the 2HDM, as was explained in ref.~\cite{Ferreira:2010bm}, they may be extended
to the Yukawa sector. The Yukawa matrices $\Gamma_i$, then, must obey the following relations,
\bea
X_\alpha \Gamma_1^\ast - (\cos \theta \Gamma_1 - \sin \theta \Gamma_2) X_\beta = 0,
\nonumber\\
X_\alpha \Gamma_2^\ast - (\sin \theta \Gamma_1 + \cos \theta \Gamma_2) X_\beta = 0\,
\label{eq:gamtr}
\eea
where the matrices $X_x$ are given by
\beq
X_x\,=\,\begin{pmatrix} \cos x & \sin x & 0 \\ -\sin x & \cos x & 0 \\
0 & 0 & 1 \end{pmatrix}
\eeq
and the angle $\theta$ (like the angles $\alpha$ and $\beta$) can be taken, without loss of generality, to be between
0 and $\pi/2$ and describes each possible CP symmetry: $\theta = 0$ corresponds to CP1; $\theta = \pi/2$ to CP2; and any
arbitrary angle $0 < \theta < \pi/2$ yields CP3. An analogous equation to~\eqref{eq:gamtr} (with different, independent
angles $\gamma$ replacing $\beta$) is valid for the up quark Yukawa matrices $\Delta_i$. Solving ~\eqref{eq:gamtr} for
CP2 and CP3 one then finds:
\begin{itemize}
\item For the CP2 symmetry, eq.~\eqref{eq:gamtr} is satisfied (for $\theta = \pi/2$ and $\alpha =\beta = \pi/4$) by
$\Gamma$ matrices of the form~\cite{Ferreira:2010bm}
\beq
\Gamma_1 \,=\,\begin{pmatrix} a_{11} & a_{12} & 0 \\ a_{12} & -a_{11} & 0 \\
0 & 0 & 0 \end{pmatrix}\;\;,\;\;
\Gamma_2 \,=\,\begin{pmatrix} -a_{12}^* & a_{11}^* & 0 \\ a_{11}^* & a_{12}^* & 0 \\
0 & 0 & 0 \end{pmatrix}\;.
\label{eq:yupcp2}
\eeq
Analogous expressions are then found for the $\Delta$ and $\Pi$ matrices, with different coefficients $b_{ij}$
and $c_{ij}$ instead of $a_{ij}$. As is plain to see, these matrices imply that one of the up and down quarks and a charged lepton
will be massless.
\item For the CP3 symmetry, eq.~\eqref{eq:gamtr} is satisfied (for $\theta = \alpha =\beta = \pi/3$) by
$\Gamma$ matrices of the form~\cite{Ferreira:2010bm}
\beq
\Gamma_1 \,=\,\begin{pmatrix} i\,a_{11} & i\,a_{12} & a_{13} \\ i\,a_{12} & -i\,a_{11} & a_{23} \\
a_{31} & a_{32} & 0 \end{pmatrix}\;\;,\;\;
\Gamma_2 \,=\,\begin{pmatrix} i\,a_{12} & -i\,a_{11} & -a_{23} \\ -i\,a_{11} & -i\,a_{12} & a_{13} \\
-a_{32} & a_{31} & 0 \end{pmatrix}\,,
\label{eq:yupcp3}
\eeq
with the $a_{ij}$ real.
Analogous matrices are then found for the $\Delta$ and $\Pi$ matrices, with different coefficients
$b_{ij}$ and $c_{ij}$ instead of $a_{ij}$. These matrices yield three generations of massive charged
fermions, and it was possible to perform a numerical fit reproducing the known quark and lepton masses;
however, that fit could not reproduce the value of the Jarlskog invariant\footnote{Notice, however,
that the phenomenological problems with the CP2 and CP3 Yukawa sector may be solved by adding vector-like
fermions to the model~\cite{Draper:2016cag,Draper:2020tyq}.}.
\end{itemize}

With Yukawa matrices that comply with symmetries CP2 and CP3 -- and therefore, as has been explained, fermionic
contributions to RG running will respect the relations between scalar quartic couplings for those models --
we can verify whether the unusual $m^2_{11} + m^2_{22} = 0$ relation is also preserved when one includes
the fermion sector in the model. Let us show how this works explicitly at one-loop -- the fermionic contributions to the
$\beta$-functions of $m^2_{11}$ and $m^2_{22}$ in eqs.~\eqref{eq:betam} are given,
for the most general 2HDM, by (see, 
for instance,~\cite{Staub:2008uz,Staub:2009bi,Staub:2010jh,Staub:2012pb,Staub:2013tta,Lyonnet:2016xiz}):
\bea
\beta^{F,1L}_{m_{11}^2} &=& \left[3 \,\textrm{Tr}(\Delta_1 \Delta_1^\dagger) \,+\,
3 \,\textrm{Tr}(\Gamma_1 \Gamma_1^\dagger)\,+\,  \textrm{Tr}(\Pi_1 \Pi_1^\dagger)\right]\,m^2_{11}
\nonumber \\
& & - \left\{ \left[3 \,\textrm{Tr}(\Delta_1^\dagger \Delta_2) \,+\, 3 \,\textrm{Tr}(\Gamma_1^\dagger \Gamma_2)
\,+\, \textrm{Tr}(\Pi_1^\dagger \Pi_2) \right]\,m^2_{12}\,+\, \textrm{h.c.} \right\}
\,, \nonumber \\
\beta^{F,1L}_{m_{22}^2} &=& \left[3 \textrm{Tr}(\Delta_2 \Delta_2^\dagger) \,+\,
3 \textrm{Tr}(\Gamma_2 \Gamma_2^\dagger)\,+\,  \textrm{Tr}(\Pi_2 \Pi_2^\dagger)\right]\,m^2_{22}
\nonumber \\
& & - \left\{ \left[3 \,\textrm{Tr}(\Delta_1^\dagger \Delta_2) \,+\, 3 \,\textrm{Tr}(\Gamma_1^\dagger \Gamma_2)
\,+\, \textrm{Tr}(\Pi_1^\dagger \Pi_2) \right]\,m^2_{12}\,+\, \textrm{h.c.} \right\}
\,,
\label{eq:betamf}
\eea
We then see something remarkable -- for both the CP2 or CP3 Yukawa textures (eqs.~\eqref{eq:yupcp2}
and~\eqref{eq:yupcp3} respectively),  one obtains
\beq
\textrm{Tr}(\Delta_1 \Delta_1^\dagger) = \textrm{Tr}(\Delta_2 \Delta_2^\dagger)\;\;,\;\;
\textrm{Tr}(\Gamma_1 \Gamma_1^\dagger) = \textrm{Tr}(\Gamma_2 \Gamma_2^\dagger)\;\;,\;\;
\textrm{Tr}(\Pi_1 \Pi_1^\dagger) = \textrm{Tr}(\Pi_2 \Pi_2^\dagger)\,,
\label{eq:y2}
\eeq
as well as
\beq
\textrm{Tr}(\Delta_1 \Delta_2^\dagger) = \textrm{Tr}(\Gamma_1 \Gamma_2^\dagger)\;=\; \textrm{Tr}(\Pi_1 \Pi_2^\dagger) \;=\; 0.
\label{eq:yiyi}
\eeq
Hence,
\beq
\beta^{F,1L}_{m_{11}^2 + m_{22}^2} \,=\,\left[3 \,\textrm{Tr}(\Delta_1 \Delta_1^\dagger) \,+\,
3 \,\textrm{Tr}(\Gamma_1 \Gamma_1^\dagger)\,+\,  \textrm{Tr}(\Pi_1 \Pi_1^\dagger)\right]\,\left( m^2_{11} + m_{22}^2\right)
\label{eq:yiyj}
\eeq
and therefore it has been shown that the $m^2_{11} + m_{22}^2 = 0$ condition is preserved under RGE running for CP2 and CP3 invariant theories at
one-loop, even including the fermionic sector.

An all-order result is beyond our skills, but we can at least extend this demonstration to two loops.
We can use the SARAH~\cite{Staub:2008uz,Staub:2009bi,Staub:2010jh,Staub:2012pb,Staub:2013tta} package and adapt its results
for a 2HDM Type-III
model\footnote{The reader should be aware that, up to version 4.15.1 of SARAH, there is a bug in the code concerning non-supersymmetric
beta-functions for the squared mass couplings in  model III. This arises when Yukawa couplings induce mixing between the doublets. The issue
has been identified and a patch is available from SARAH's keepers. Many thanks to Mark Goodsell for his help in this matter.}
for the specific Yukawa matrices of eqs.~\eqref{eq:yupcp2} and~\eqref{eq:yupcp3}. Doing so, we find that when the Yukawa
matrices have the CP2/CP3 structures and the potential obeys the 0CP2/0CP3 symmetry, the two-loop beta functions for
the quadratic scalar couplings, including fermions, satisfy
\beq
\beta^{2L}_{m_{11}^2 + m_{22}^2} \,=\,X\, (m_{11}^2 + m_{22}^2)\,,
\eeq
where the quantity ``X" contains contributions from all dimensionless couplings of the model (scalar, gauge, Yukawa).
And therefore, as in the one-loop case, we verify that $m^2_{11} + m_{22}^2 = 0$ is preserved by RG running
up to two loops at least. This may also be independently verified using the PyR@TE package~\cite{Lyonnet:2016xiz}.
Wishing to go beyond the ``black box" of these remarkable packages, we performed a simplified verification of these results
to understand the cancellations between different terms necessary for them to occur, and show it, as a curiosity, in
Appendix~\ref{ap:ferm}.

It seems therefore likely that invariance under the $r_0$ symmetry, at least for the 0CP2/0CP3 versions,
can be extended to three-loop order in the Yukawa sector --
or indeed to all orders, as we argued was the case for the scalar and gauge contributions.

\section{Conclusions}
\label{sec:conclusions}

We found a set of constraints on 2HDM scalar parameters which is RG invariant to all orders when one considers only scalar and gauge
interactions -- and which can be invariant to at least two loops if fermions are also included. To do so we analysed
the beta-functions of the parameters of the model and discovered fixed points -- valid to all orders in scalar and
gauge couplings -- which correspond to relations between 2HDM parameters which do not coincide with any of
the known six symmetries of the $SU(2)\times U(1)$ scalar potential. Those relations, given by what we called
the $r_0$-symmetry\footnote{Considering the names of the authors, the only other reasonable nomenclature
would be the GOOF symmetry, and we do not think such possibility would be well-met in the community.}, are
\beq
m^2_{11}\,+\,m^2_{22}\,=\,0\;\;\; , \;\;\; \lambda_1 = \lambda_2 \;\;,\;\; \lambda_6 = -\lambda_7\,,
\eeq
which have also been shown to be basis-invariant.

It is well known that invariance of a system under a symmetry imposes certain relations among its parameters; and
those relations will be preserved under renormalization to all orders, constituting fixed points of its RG equations.
What we have found here for the 2HDM is in some sense the opposite situation: we found fixed points of the RG equations,
valid to all orders of perturbation theory, but do not know what symmetry operation upon the model's fields may
cause them to appear. We showed that one way of understanding the relations obtained for the parameters of the scalar
potential is to consider an ``$r_0$ sign change" in the gauge invariant bilinear $r_0$ -- but though that is helpful
as a formal way of understanding our results, it raises serious questions, as the scalar kinetic terms are not invariant under
the transformation $r_0 \rightarrow - r_0$. Further, this transformation is impossible to obtain via unitary or antiunitary transformations
on the doublets. We propose a (very strange) set of transformations on the scalar and gauge fields in Appendix~\ref{ap:comp},
but it is unclear whether or not it constitutes a mathematical trick only.

Therefore, strictly speaking, we may not have identified ``symmetries" of the 2HDM. If the reader wishes, call them instead
``{\em relations between 2HDM parameters which yield fixed points of the RG equations to all orders of perturbation
theory and are therefore preserved under renormalization}". But given that the several models we discuss here will benefit
from all features of symmetries when these all-order invariant relations are considered, we believe that calling these ``symmetries"
is justified, and challenge our colleagues to find the field transformations which yield them.

Combining the $r_0$-symmetry with the other known six symmetries yields seven new symmetry classes. We briefly investigated the
phenomenological aspects of each of those symmetries, considering possible soft breaking terms. The impact of the new symmetries
on physical parameters -- masses, couplings of scalar-gauge boson and scalar self interactions -- was shown, with simple
relations between those observables obtained. We also concluded that the
$r_0$-symmetry has measurable impacts on the 2HDM, namely it prevents the existence of a {\em decoupling limit} -- the
$r_0$-symmetry, coupled with minimization conditions, eliminates all dependence on squared mass parameters in the
scalars' physical masses, and therefore, in models invariant under the $r_0$-symmetry, the non-SM-like scalars
cannot be arbitrarily heavy. We found bounds of a little above 700 GeV for both charged and neutral scalars. Therefore
these models can easily be disproven experimentally, if bounds on extra scalar masses are found to be well above
$\sim$700 GeV when new LHC data is analysed.

Another possibly interesting phenomenological consequence of the $r_0$-symmetry occurs for the softly broken 0U(1) model,
where the extra CP-even scalar and the CP-odd one were found to be mass degenerate at tree level. However, since the CP-even
particle has different interactions than the CP-odd one (it couples to $W$ and $Z$ pairs, for instance, as well as to
charged scalar pairs, $H^+ H^-$), this mass degeneracy will be lifted via radiative corrections. A full one-loop
calculation is necessary to determine the mass splitting between these two scalars, but one might expect that it will not
be sizeable. Hence, if a CP-even scalar and a CP-odd one were discovered at the LHC with a small mass difference
between them, the $r_0$-symmetry coupled with a Peccei-Quinn U(1) symmetry may therefore provide a simple, natural way
to explain it.

We have found several instances where the $r_0$-symmetry prevents tree-level spontaneous symmetry breaking -- in the
0CP1 model, spontaneous CP violation is found to require, at tree-level, an RG-unstable relation among quartic couplings;
likewise, in the 0Z2 model, a vacuum where both doublets acquire vevs and spontaneous breaking of the $Z_2$ symmetry
would occur, is found to also require an RG-unstable relation among quartic couplings (albeit a different one); and
the same occurs for spontaneous U(1) breaking in the 0U(1) model, with yet another RG-unstable condition on the quartic couplings
necessary for the tree-level minimization equations to have a solution. Further, in these cases, it was found that the
tree-level minimization equations did not allow for the unequivocal determination of the doublets' vevs. These are situations
where a one-loop minimization is necessary, to verify whether radiative corrections allow the spontaneous breaking of
these symmetries, as in the Georgi-Pais mechanism~\cite{Georgi:1974au}.

We showed that, at least for some of the models proposed, it is possible to extend the $r_0$-symmetry to the full lagrangian,
including fermions. We did not obtain an all-order result, but were capable of showing that, at least up to two loops,
the 0CP2 symmetry, including CP-symmetric Yukawa matrices, was a symmetry of the full larangian. Likewise, the 0CP3 model,
with CP3-symmetric Yukawa matrices, is fully consistent up to two-loops in the fermionic sector, at least. This strongly
suggests that these parameter relations may indeed be preserved under renormalization to all orders of perturbation theory,
even including Yukawa interactions. The CP2 and CP3 Yukawas considered were just a ``case study" to prove extension to fermions
of the  $r_0$-symmetry was possible,  but they are not necessarily the only ones -- others may be found. The CP2 an CP3 Yukawa
textures have phenomenological problems associated with them (massless fermions in the former case; wrong values for the
Jarlskog invariant for the latter), which may be solved by enlarging the particle content of the 2HDM via the introduction of
vector-like fermions~\cite{Draper:2016cag,Draper:2020tyq}. It would be interesting to verify whether, with the extra fermion
content, it would still be possible for the 0CP2 and 0CP3 models to be RG-invariant (at least to two loops) when including
the Yukawa sector.

In conclusion, we have shown that the 2HDM includes regions of parameter space were relations between scalar couplings
are RG-invariant to all orders -- and for at least two cases, at least to two-loop order when one includes Yukawa interactions.
The models boasting the new  $r_0$-symmetry have interesting phenomenology and leave plenty of questions upon for future
avenues of research, the more pressing one of which may well be whether there are transformations on the
fields of the model which reproduce the $r_0$-symmetry. Appendix~\ref{ap:comp} has one such proposal which
works mathematically, but whose physical meaning is unclear.

\section*{Acknowledgements}
P.M.F. is supported
by \textit{Funda\c c\~ao para a Ci\^encia e a Tecnologia} (FCT)
through contracts
UIDB/00618/2020, UIDP/00618/2020, CERN/FIS-PAR/0004/2019, CERN/FIS-PAR/0014/2019 and CERN/FIS-PAR/ 0025/2021.
The work of B.G. is supported in part by the National Science Centre (Poland) as a research project  2020/37/B/ST2/02746.
The research of P.O. has been supported in part by the Research Council of Norway. Many thanks to Mark Goodsell
for providing the SARAH patch required to obtain the correct results for Model III, and for assistance
with PyR@TE.

\appendix

\section{Imaginary spacetime - a proposed transformation originating the $r_0$ symmetry}
\label{ap:comp}

In section~\ref{sec:r0} we showed that the $r_0$ symmetry could be interpreted as a change in sign
in the $r_0$ bilinear. However, though this formally worked for the scalar potential, the transformation
$r_0 \rightarrow -r_0$ could not be extended to the theory's kinetic terms in any obvious manner.
In this appendix we will show a curiosity: it is possible to obtain a transformation of fields and
spacetime coordinates which leave the lagrangian invariant under the $r_0$ symmetry, but such
a transformation involves a complex spacetime and gauge-breaking relations between the fields, though
the final theory is gauge-invariant.

Let us for this purpose parameterize the doublets as
\bea
\Phi_1=
\begin{pmatrix}
	\phi_1+i\phi_2 \\
	\phi_3+i\phi_4
\end{pmatrix}, \quad
\Phi_2=
\begin{pmatrix}
	\phi_5+i\phi_6 \\
	\phi_7+i\phi_8
\end{pmatrix},
\eea
with all fields $\phi_i$ Hermitian.
We find that the bilinears can be expressed as
\bea
r_0&=&\frac{1}{2}(\phi_1^2+\phi_2^2+\phi_3^2+\phi_4^2+\phi_5^2+\phi_6^2+\phi_7^2+\phi_8^2),\nonumber\\
r_1&=&\phi_1\phi_5+\phi_2\phi_6+\phi_3\phi_7+\phi_4\phi_8,\nonumber\\
r_2&=&-\phi_2\phi_5+\phi_1\phi_6-\phi_4\phi_7+\phi_3\phi_8,\nonumber\\
r_3&=&\frac{1}{2}(\phi_1^2+\phi_2^2+\phi_3^2+\phi_4^2-\phi_5^2-\phi_6^2-\phi_7^2-\phi_8^2).
\eea
We are looking for a transformation that makes $r_0$ change sign, while $r_1$, $r_2$ and $r_3$ are unchanged
-- this, indeed, is the interpretation we made of the $r_0$ symmetry in section~\ref{sec:r0}.
The transformation
\bea
\begin{pmatrix}
	\phi_1\\
	\phi_2\\
	\phi_3\\
	\phi_4\\
	\phi_5\\
	\phi_6\\
	\phi_7\\
	\phi_8
\end{pmatrix}
\to
\begin{pmatrix}
	0 & 0 & 0 & 0 & 0 & i & 0 & 0 \\
	0 & 0 & 0 & 0 & i & 0 & 0 & 0 \\
	0 & 0 & 0 & 0 & 0 & 0 & 0 & i \\
	0 & 0 & 0 & 0 & 0 & 0 & i & 0 \\
	0 & -i & 0 & 0 & 0 & 0 & 0 & 0 \\
	-i & 0 & 0 & 0 & 0 & 0 & 0 & 0 \\
	0 & 0 & 0 & -i & 0 & 0 & 0 & 0 \\
	0 & 0 & -i & 0 & 0 & 0 & 0 & 0
\end{pmatrix}
\begin{pmatrix}
	\phi_1\\
	\phi_2\\
	\phi_3\\
	\phi_4\\
	\phi_5\\
	\phi_6\\
	\phi_7\\
	\phi_8
\end{pmatrix}\label{transform}
\eea
accomplishes this. This transformation implies
\begin{alignat}{2}
\Phi_1&\to-\Phi_2^* &\quad
\Phi_1^\dagger&\to \Phi_2^T,\nonumber\\
\Phi_2&\to\Phi_1^*, &\quad
\Phi_2^\dagger&\to -\Phi_1^T.
\label{eq:trandou}
\end{alignat}
Notice that this transformation, applied to the real component fields of the doublets, forces
each doublet and their hermitian conjugates to transform differently than they should. Indeed,
the transformation of $\Phi_1^\dagger$ above is {\em not} the hermitian conjugate of the transformation
of $\Phi_1$, and the same holds for the second doublet. This suggests that behind the $r_0$ symmetry is
a type of formalism in which $\Phi_i$ and $\Phi_i^\dagger$ should be treated as independent objects. Notice,
too, that the transformation of eq.~\eqref{transform} is akin to a $Z_4$ symmetry in the sense that, to recover
the original doublets, one needs to apply it {\em four} times. This is more easily seen from eq.~\eqref{eq:trandou}.


Let us now verify whether one can make the scalar kinetic terms invariant under the transformation of eq.~\eqref{transform}.
The scalar covariant derivatives are given by
\begin{equation}
D^\mu=\partial^\mu+\frac{ig}{2}\sigma_iW_i^\mu+i\frac{g^\prime}{2}B^\mu,
\end{equation}
so that
the scalar kinetic part of the Lagrangian can be written as
\begin{equation} \label{Eq:gauge-ISS}
	\lcal_k=(D_\mu \Phi_1)^\dagger(D^\mu \Phi_1) + (D_\mu \Phi_2)^\dagger(D^\mu \Phi_2).
\end{equation}
The kinetic terms are invariant under the transformations of eq.~\eqref{transform}
if we combine them with
\begin{gather}
\partial_\mu \to -i\partial_\mu,\nonumber\\
B_\mu \to i B_\mu, \nonumber\\
W_{1\mu} \to i W_{1\mu},\quad
W_{2\mu} \to -i W_{2\mu},\quad
W_{3\mu} \to i W_{3\mu}.
\label{eq:igau}
\end{gather}
We shall call this the extended $r_0$ transformation. Notice how the first of these corresponds to a transformation
on the spacetime coordinates themselves,
\beq
x_\mu \to i x_\mu \,.
\eeq
Strange as this transformation is, we observe that it leaves the spacetime integration $d^4 x$ invariant.

The imaginary transformations on the gauge fields of eq.~\eqref{eq:igau} would correspond to, for the gauge mass eigenstates,
\begin{gather}
A_\mu \to i A_\mu,\nonumber\\
Z_\mu \to i Z_\mu,\nonumber\\
W_\mu \to i W_\mu^+,\quad
W_\mu^+ \to i W_\mu.
\end{gather}
where $W_1^\mu=\frac{1}{\sqrt{2}}(W^{+\mu}+W^{-\mu})$, $W_2^\mu=\frac{i}{\sqrt{2}}(W^{+\mu}-W^{-\mu})$, $W_3^\mu=\cos\thetaW Z^\mu+\sin\thetaW A^\mu$ and $B^\mu=-\sin\thetaW Z^\mu+\cos\thetaW A^\mu$~\footnote{These are oddly consistent. It is curious to observe that, for the simple
case in electromagnetism of the 4-potential produced by a moving point charge, if one makes $x_\mu \to i x_\mu$, we indeed obtain in
that situation $A_\mu \to i A_\mu$.}.

The net effect of the extended $r_0$ transformation is that the covariant derivatives acting on the doublets transform according to
\begin{align}
D^\mu\Phi_1 &\rightarrow  i\left(D^\mu\Phi_2\right)^*, \quad\;\; \left(D^\mu\Phi_1\right)^\dagger \rightarrow -i\left(D^\mu\Phi_2\right)^T,\nonumber\\
D^\mu\Phi_2 &\rightarrow  -i\left(D^\mu\Phi_1\right)^*,\quad \left(D^\mu\Phi_2\right)^\dagger \rightarrow i\left(D^\mu\Phi_1\right)^T
\end{align}
and then it is easy to see that the scalar kinetic terms are clearly invariant under these transformations.

Having found a gauge field transformation necessary to render invariant the scalar kinetic terms,
we must then worry about the gauge kinetic terms themselves. These can be written compactly as
\bea
\lcal^B&=&-\frac{1}{4}B_{\mu\nu}B^{\mu\nu}-\frac{1}{4}W_{i\mu\nu}W_i^{\mu\nu},
\eea
where $B^{\mu\nu}=\partial^\nu B^\mu-\partial^\mu B^\nu$ and $W_i^{\mu\nu}=\partial^\nu W_i^\mu-\partial^\mu W_i^\nu+g\epsilon_{ijk}W_j^\mu W_k^\nu$.
We find that under the extended $r_0$ transformation defined above in eq.~\eqref{eq:igau} we have
\begin{gather}
B^{\mu \nu} \to  B^{\mu \nu},\nonumber\\
W_1^{\mu \nu}  \to  W_1^{\mu \nu},\quad
W_2^{\mu \nu}  \to  -W_2^{\mu \nu},\quad
W_3^{\mu \nu} \to W_3^{\mu \nu},
\end{gather}
and it is then  clear that $\lcal^B$ is invariant under the extended $r_0$ transformation. A generalization of these
imaginary transformations to fermionic fields should also be possible.

\section{Two loop fermionic beta-functions and the condition $m^2_{11} + m^2_{22} = 0$}
\label{ap:ferm}

For our purposes -- demonstrating that the $m^2_{11} + m_{22}^2 = 0$ condition is left invariant under
RG running at two loops by CP2 and/or CP3 Yukawa matrices\footnote{To be precise, we should add that the conditions
$\lambda_1 = \lambda_2$ and $\lambda_6 = -\lambda_7$ are also left invariant under RG running, but we already know
that that is the case
for both CP2 and CP3 symmetries (with $\lambda_6 = \lambda_7 = 0$ in the latter case).} -- we do not need the {\em exact}
form of the two-loop beta functions. All we need to do is analyse the structure
 of Yukawa couplings emerging from all contributions to the beta functions and deduce that they are such that $\beta_{m^2_{11} + m^2_{22}}$
 ends up being proportional to $m^2_{11} + m^2_{22}$. Let us show, through
a partial calculation, how this comes about.

\begin{figure}[t]
\begin{center}
\includegraphics[height=4cm,angle=0]{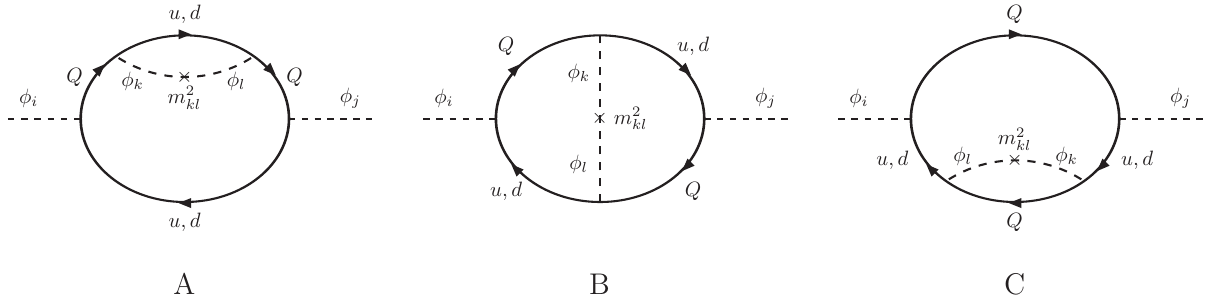}
\end{center}
\vspace{-6mm}
\caption{Feynman diagrams contributing to the beta functions for the quadratic scalar coefficients involving only
Yukawa interactions. The ``$\times$" symbol
denotes a ``mass insertion" corresponding to the $m^2_{ij}$ coefficients.
}
\label{fig:diag}
\end{figure}

\begin{itemize}
\item Yukawa-only contributions
\end{itemize}
Using dimensional regularization, wherein the spacetime dimension is taken to be $4 - 2\epsilon$ with $\epsilon \rightarrow 0$,
the two-loop beta functions for the scalar quadratic coefficients arise from Feynman diagrams such as those shown in
figure~\ref{fig:diag}, where we only considered contributions from quark interactions\footnote{The contributions from leptons would be
simpler as we are not considering Dirac neutrino masses, and the argument would follow in the same way.}. Specifically, each Feynman
diagram will have a pole in $1/\epsilon$, the coefficient of which is the contribution to the beta function.
In figure~\ref{fig:diag} the cross, ``$\times$", can be thought of as a  ``mass insertion", or rather as a ``vertex" where a
$m^2_{11}$ coefficient corresponds to a continuous $\Phi_1$ line;
a $m^2_{22}$ coefficient corresponds to a continuous $\Phi_2$ line; and a $-m^2_{12}$ ($-m^2_{21} = -{m}_{12}^{2*}$) coefficient
turns a $\Phi_1$ ($\Phi_2$) line into a $\Phi_2$ ($\Phi_1$) one. Note that one should also consider a diagram analogous to
``B" but with the fermionic lines flowing in the opposite way -- the diagram topology of both possibilities is the same,
but the Yukawa structures arising from each differ.

The crucial point for the argument that follows is that each of
the diagrams of figure~\ref{fig:diag} has the same topology, and will therefore yield a given prefactor -- $A$, $B$ or $C$ --
containing the same pole coefficient, symmetry factor, sum on colour indices, etc, whatever the specific Yukawa couplings or $m^2_{ij}$ coefficients one
may consider for that topology. As an example, consider diagram $A$. If one takes $m^2_{kl} = m^2_{11}$, for instance, one is left
with four possibilities for the fermion lines: (i) only ``$u$" lines, (ii) only ``$d$" lines, (iii) upper ``$u$ and lower ``$d$"
lines and (iv) vice-versa. These will correspond to combinations of Yukawa matrices given by, respectively,
$\textrm{Tr}(\Delta_1 \Delta_1^\dagger\Delta_1 \Delta_1^\dagger)$, $\textrm{Tr}(\Gamma_1 \Gamma_1^\dagger\Gamma_1 \Gamma_1^\dagger)$,
$\textrm{Tr}(\Gamma_1 \Delta_1^\dagger\Delta_1 \Gamma_1^\dagger)$ and $\textrm{Tr}(\Delta_1 \Gamma_1^\dagger\Gamma_1 \Delta_1^\dagger)$,
but all will be multiplied by $m^2_{11}$ {\em and} the same factor $A$, characteristic of this specific diagram topology.
In this way, it is a simple exercise to write down the contributions from the ``A" diagram to $\beta^{F,2L}_{m_{11}^2}$, obtaining
\bea
\beta^{F(A),2L}_{m_{11}^2} &=& A \left[\textrm{Tr}(\Delta_1 \Delta_1^\dagger\Delta_1 \Delta_1^\dagger) +
\textrm{Tr}(\Gamma_1 \Gamma_1^\dagger\Gamma_1 \Gamma_1^\dagger) + \textrm{Tr}(\Gamma_1 \Delta_1^\dagger\Delta_1 \Gamma_1^\dagger)
 + \textrm{Tr}(\Delta_1 \Gamma_1^\dagger\Gamma_1 \Delta_1^\dagger)\right] m^2_{11}  \nonumber \\
 & &  +A \left[\textrm{Tr}(\Delta_1 \Delta_2^\dagger\Delta_2 \Delta_1^\dagger) +
\textrm{Tr}(\Gamma_1 \Gamma_2^\dagger\Gamma_2 \Gamma_1^\dagger) + \textrm{Tr}(\Gamma_1 \Delta_2^\dagger\Delta_2 \Gamma_1^\dagger)
 + \textrm{Tr}(\Delta_1 \Gamma_2^\dagger\Gamma_2 \Delta_1^\dagger) \right] m^2_{22}  \nonumber \\
 & & -A \left\{ \frac{}{} \left[\textrm{Tr}(\Delta_1 \Delta_1^\dagger\Delta_2 \Delta_1^\dagger) +
\textrm{Tr}(\Gamma_1 \Gamma_1^\dagger\Gamma_2 \Gamma_1^\dagger) + \textrm{Tr}(\Gamma_1 \Delta_1^\dagger\Delta_2 \Gamma_1^\dagger)
 + \textrm{Tr}(\Delta_1 \Gamma_1^\dagger\Gamma_2 \Delta_1^\dagger) \right] m^2_{12} \right. \nonumber \\
 & & \;\;\;\;\;+ \,\left. \frac{}{} \textrm{h.c.} \right\}\;.
 \label{eq:A11}
\eea
It is then trivial to obtain the contributions from the ``A" diagram to $\beta^{F,2L}_{m_{22}^2}$, by taking the result above and
performing the exchange $1\leftrightarrow 2$ throughout, which results in
\bea
\beta^{F(A),2L}_{m_{22}^2} &=& A \left[\textrm{Tr}(\Delta_2 \Delta_1^\dagger\Delta_1 \Delta_2^\dagger) +
\textrm{Tr}(\Gamma_2 \Gamma_1^\dagger\Gamma_1 \Gamma_2^\dagger) + \textrm{Tr}(\Gamma_2 \Delta_1^\dagger\Delta_1 \Gamma_2^\dagger)
 + \textrm{Tr}(\Delta_2 \Gamma_1^\dagger\Gamma_1 \Delta_2^\dagger) \right] m^2_{11}  \nonumber \\
& & +A \left[\textrm{Tr}(\Delta_2 \Delta_2^\dagger\Delta_2 \Delta_2^\dagger) +
\textrm{Tr}(\Gamma_2 \Gamma_2^\dagger\Gamma_2 \Gamma_2^\dagger) + \textrm{Tr}(\Gamma_2 \Delta_2^\dagger\Delta_2 \Gamma_2^\dagger)
 + \textrm{Tr}(\Delta_2 \Gamma_2^\dagger\Gamma_2 \Delta_2^\dagger)\right] m^2_{22}  \nonumber \\
 & & -A \left\{ \frac{}{} \left[\textrm{Tr}(\Delta_2 \Delta_1^\dagger\Delta_2 \Delta_2^\dagger) +
\textrm{Tr}(\Gamma_2 \Gamma_1^\dagger\Gamma_2 \Gamma_2^\dagger) + \textrm{Tr}(\Gamma_2 \Delta_1^\dagger\Delta_2 \Gamma_2^\dagger)
 + \textrm{Tr}(\Delta_2 \Gamma_1^\dagger\Gamma_2 \Delta_2^\dagger) \right] m^2_{12} \right. \nonumber \\
 & & \;\;\;\;\;+ \,\left. \frac{}{} \textrm{h.c.} \right\}\;.
  \label{eq:A22}
\eea
At this stage a direct calculation with the Yukawa structures from CP2 (eq.~\eqref{eq:yupcp2})
or CP3 (eq.~\eqref{eq:yupcp3}) shows that:
\begin{itemize}
\item The quantity in square brackets multiplying $m^2_{11}$ in eq.~\eqref{eq:A11} is equal to
the quantity in square brackets multiplying $m^2_{22}$ in eq.~\eqref{eq:A22}.
\item The quantity in square brackets multiplying $m^2_{22}$ in eq.~\eqref{eq:A11} is equal to
the quantity in square brackets multiplying $m^2_{11}$ in eq.~\eqref{eq:A22} -- this equality may be seen even
without using the specific CP2 or CP3 Yukawa structures, it is a direct consequence of the cyclical
property of the trace of matrix products.
\item If one sums eqs.~\eqref{eq:A11} and~\eqref{eq:A22}, the terms proportional to $m^2_{12}$ cancel out -- in fact,
for the CP3 Yukawa textures, those terms are individually zero for each of the equations mentioned.
\end{itemize}
As a result, we obtain
\bea
\beta^{F(A),2L}_{m_{11}^2 + m_{22}^2} & = & A \left[
\textrm{Tr}(\Delta_1 \Delta_1^\dagger\Delta_1 \Delta_1^\dagger) +
\textrm{Tr}(\Gamma_1 \Gamma_1^\dagger\Gamma_1 \Gamma_1^\dagger) +
\textrm{Tr}(\Gamma_1 \Delta_1^\dagger\Delta_1 \Gamma_1^\dagger) \right. \nonumber \\
 & &  +\textrm{Tr}(\Delta_1 \Gamma_1^\dagger\Gamma_1 \Delta_1^\dagger)+
\textrm{Tr}(\Delta_2 \Delta_1^\dagger\Delta_1 \Delta_2^\dagger) +
\textrm{Tr}(\Gamma_2 \Gamma_1^\dagger\Gamma_1 \Gamma_2^\dagger)  \nonumber \\
 & & \left. +\textrm{Tr}(\Gamma_2 \Delta_1^\dagger\Delta_1 \Gamma_2^\dagger) +
\textrm{Tr}(\Delta_2 \Gamma_1^\dagger\Gamma_1 \Delta_2^\dagger)
 \right]\,\left( m^2_{11} + m_{22}^2\right)
\eea
and again, as in eq.~\eqref{eq:yiyj}, we see the proportionality to $(m^2_{11} + m_{22}^2)$.

The same thing happens for the other diagrams from fig.~\ref{fig:diag}. For diagram ``B", for instance -- and summing both possibilities
of direction of fermionic lines, as mentioned above --, one has
\bea
\beta^{F(B),2L}_{m_{11}^2} &=& 2 B \left[\textrm{Tr}(\Delta_1 \Delta_1^\dagger\Delta_1 \Delta_1^\dagger) +
\textrm{Tr}(\Gamma_1 \Gamma_1^\dagger\Gamma_1 \Gamma_1^\dagger) + \textrm{Tr}(\Gamma_1 \Delta_1^\dagger\Delta_1 \Gamma_1^\dagger)
 + \textrm{Tr}(\Delta_1 \Gamma_1^\dagger\Gamma_1 \Delta_1^\dagger)\right] m^2_{11}  \nonumber \\
 & &  +B \left[\textrm{Tr}(\Delta_1 \Delta_2^\dagger\Delta_1 \Delta_2^\dagger) +
\textrm{Tr}(\Gamma_1 \Gamma_2^\dagger\Gamma_1 \Gamma_2^\dagger) + \textrm{Tr}(\Gamma_1 \Delta_2^\dagger\Delta_1 \Gamma_2^\dagger)
 + \textrm{Tr}(\Delta_1 \Gamma_2^\dagger\Gamma_1 \Delta_2^\dagger) + \textrm{h.c.} \right] m^2_{22}
 \nonumber \\
 & & -B \left\{ \frac{}{} \left[\textrm{Tr}(\Delta_1 \Delta_1^\dagger\Delta_1 \Delta_2^\dagger) +
\textrm{Tr}(\Gamma_1 \Gamma_1^\dagger\Gamma_1 \Gamma_2^\dagger) + \textrm{Tr}(\Gamma_1 \Delta_1^\dagger\Delta_1 \Gamma_2^\dagger)
 + \textrm{Tr}(\Delta_1 \Gamma_1^\dagger\Gamma_1 \Delta_2^\dagger) \right] m^2_{12} \right. \nonumber \\
 & & \;\;\;\;\;+ \,\left. \frac{}{} \textrm{h.c.} \right\}\;.
 \label{eq:B11}
\eea
and
\bea
\beta^{F(B),2L}_{m_{22}^2} &=& B \left[\textrm{Tr}(\Delta_2 \Delta_1^\dagger\Delta_2 \Delta_1^\dagger) +
\textrm{Tr}(\Gamma_2 \Gamma_1^\dagger\Gamma_2 \Gamma_1^\dagger) + \textrm{Tr}(\Gamma_2 \Delta_1^\dagger\Delta_2 \Gamma_1^\dagger)
 + \textrm{Tr}(\Delta_2 \Gamma_1^\dagger\Gamma_2 \Delta_1^\dagger) + \textrm{h.c.}\right] m^2_{11} \nonumber \\
& & +B \left[\textrm{Tr}(\Delta_2 \Delta_2^\dagger\Delta_2 \Delta_2^\dagger) +
\textrm{Tr}(\Gamma_2 \Gamma_2^\dagger\Gamma_2 \Gamma_2^\dagger) + \textrm{Tr}(\Gamma_2 \Delta_2^\dagger\Delta_2 \Gamma_2^\dagger)
 + \textrm{Tr}(\Delta_2 \Gamma_2^\dagger\Gamma_2 \Delta_2^\dagger)\right] m^2_{22}  \nonumber \\
 & & -B \left\{ \frac{}{} \left[\textrm{Tr}(\Delta_2 \Delta_1^\dagger\Delta_2 \Delta_2^\dagger) +
\textrm{Tr}(\Gamma_2 \Gamma_1^\dagger\Gamma_2 \Gamma_2^\dagger) + \textrm{Tr}(\Gamma_2 \Delta_1^\dagger\Delta_2 \Gamma_2^\dagger)
 + \textrm{Tr}(\Delta_2 \Gamma_1^\dagger\Gamma_2 \Delta_2^\dagger) \right] m^2_{12} \right. \nonumber \\
 & & \;\;\;\;\;+ \,\left. \frac{}{} \textrm{h.c.} \right\}\;.
  \label{eq:B22}
\eea
Once more, the terms proportional to $m^2_{12}$ cancel when summing both beta functions, and one finds
\bea
\beta^{F(B),2L}_{m_{11}^2 + m_{22}^2} & = & 2 B \left[
\textrm{Tr}(\Delta_1 \Delta_1^\dagger\Delta_1 \Delta_1^\dagger) +
\textrm{Tr}(\Gamma_1 \Gamma_1^\dagger\Gamma_1 \Gamma_1^\dagger) +
\textrm{Tr}(\Gamma_1 \Delta_1^\dagger\Delta_1 \Gamma_1^\dagger)  \right. \nonumber \\
 & &  +\textrm{Tr}(\Delta_1 \Gamma_1^\dagger\Gamma_1 \Delta_1^\dagger) +
\textrm{Re} \left\{ \textrm{Tr}(\Delta_2 \Delta_1^\dagger\Delta_2 \Delta_1^\dagger) +
\textrm{Tr}(\Gamma_2 \Gamma_1^\dagger\Gamma_2 \Gamma_1^\dagger) \right. \nonumber \\
 & & \left.\left.  +\textrm{Tr}(\Gamma_2 \Delta_1^\dagger\Delta_2 \Gamma_1^\dagger)
 + \textrm{Tr}(\Delta_2 \Gamma_1^\dagger\Gamma_2 \Delta_1^\dagger)\right\}
 \right]\,\left( m^2_{11} + m_{22}^2\right)
\eea
A similar exercise may be undertaken for the diagram ``C", resulting in
\bea
\beta^{F(C),2L}_{m_{11}^2 + m_{22}^2} & = & C \left[
\textrm{Tr}(\Delta_1 \Delta_1^\dagger\Delta_1 \Delta_1^\dagger) +
\textrm{Tr}(\Gamma_1 \Gamma_1^\dagger\Gamma_1 \Gamma_1^\dagger) +
\textrm{Tr}(\Gamma_1 \Delta_1^\dagger\Delta_1 \Gamma_1^\dagger) \right. \nonumber \\
 & & +\textrm{Tr}(\Delta_1 \Gamma_1^\dagger\Gamma_1 \Delta_1^\dagger)+
\textrm{Tr}(\Delta_1 \Delta_1^\dagger\Delta_2 \Delta_2^\dagger) +
\textrm{Tr}(\Gamma_1 \Gamma_1^\dagger\Gamma_2 \Gamma_2^\dagger) \nonumber \\
 & & +\left. \textrm{Tr}(\Delta_1 \Gamma_1^\dagger\Gamma_2 \Delta_2^\dagger) +
\textrm{Tr}(\Gamma_1 \Delta_1^\dagger\Delta_2 \Gamma_2^\dagger)
 \right]\,\left( m^2_{11} + m_{22}^2\right)\,.
\eea
Finally, we verified that diagrams like those of fig.~\ref{fig:diag} with mass insertions on external
lines instead of internal ones also yield Yukawa structures such that the conclusions reached above
also hold: the beta function for $(m_{11}^2 + m_{22}^2)$ is proportional to that same quantity.

\begin{itemize}
\item Yukawa and quartic scalar coupling  contributions
\end{itemize}

\begin{figure}[h]
\centering
\includegraphics[height=4cm,angle=0]{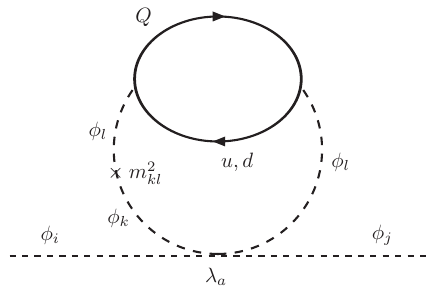}
\caption{Example of Feynman diagram contributing to the beta functions for the quadratic scalar coefficients involving
both Yukawa and scalar quartic interactions. The ``$\times$" symbol
denotes a ``mass insertion" corresponding to the $m^2_{ij}$ coefficients.
}
\label{fig:diag2}
\end{figure}

The beta functions for $m_{11}^2$ and $m_{22}^2$ also receive contributions involving Yukawa and scalar quartic
interactions, such as those exemplified in the diagram of fig.~\ref{fig:diag2}. All such contributions will be
proportional to the quadratic combinations of Yukawa couplings appearing in eqs.~\eqref{eq:y2} and~\eqref{eq:yiyi}, multiplied by a
$\lambda_a$ quartic coupling and a common factor ``$D$", in which we include the coefficient of the pole in $1/\epsilon$ and diagram
symmetry and colour factors. As an example, consider the contribution to $\beta_{m_{11}^2}$ from this diagram which is proportional to
$\lambda_1$: there will be a ``mass insertion" $m^2_{11}$, which necessitates Yukawa interactions such as $\Gamma_1 \Gamma_1^\dagger$,
which preserve the scalar doublet index, and another ``mass insertion" $m^2_{12}$, for which the Yukawa interactions must swap the doublet index from 1 to 2. This leads to
\bea
\beta_{m_{11}^2} &=&\ldots\, +\,D\,\lambda_1 \left(\frac{}{}\left[3 \,\textrm{Tr}(\Delta_1 \Delta_1^\dagger) \,+\,
3 \,\textrm{Tr}(\Gamma_1 \Gamma_1^\dagger)\,+\,  \textrm{Tr}(\Pi_1 \Pi_1^\dagger)\right]\,m^2_{11}\right.
\nonumber \\
 & & \quad\quad\quad\quad\quad\left. -\,\left\{ \left[3 \,\textrm{Tr}(\Delta_1 \Delta_2^\dagger) \,+\, 3 \,\textrm{Tr}(\Gamma_1 \Gamma_2^\dagger)
\,+\, \textrm{Tr}(\Pi_1 \Pi_2^\dagger) \right]\,m^2_{12}\,+\, \textrm{h.c.} \right\}\frac{}{}\right)\,.
\eea
Analogously, $\beta_{m_{22}^2}$  will have a term proportional to $\lambda_2$, given by
\bea
\beta_{m_{22}^2} &=&\ldots\, +\,D\,\lambda_2 \left(\frac{}{}\left[3 \,\textrm{Tr}(\Delta_2 \Delta_2^\dagger) \,+\,
3 \,\textrm{Tr}(\Gamma_2 \Gamma2^\dagger)\,+\,  \textrm{Tr}(\Pi_2 \Pi_2^\dagger)\right]\,m^2_{22}\right.
\nonumber \\
 & & \quad\quad\quad\quad\quad\left. -\,\left\{ \left[3 \,\textrm{Tr}(\Delta_2 \Delta_1^\dagger) \,+\, 3 \,\textrm{Tr}(\Gamma_2 \Gamma_1^\dagger)
\,+\, \textrm{Tr}(\Pi_2 \Pi_1^\dagger) \right]\,m^2_{12}\,+\, \textrm{h.c.} \right\}\frac{}{}\right)\,.
\eea
Given the results of eq.~\eqref{eq:yiyi}, the terms proportional to $m^2_{12}$ vanish; and since for the $r_0$ symmetry one must
have $\lambda_1 = \lambda_2$, given the results from eq.~\eqref{eq:y2} we see that once more the sum of these two contributions
yields $\beta_{m_{11}^2 + m_{22}^2} \propto m_{11}^2 + m_{22}^2$.

Much in the same manner, it is simple to obtain the terms proportional to $\lambda_3$,
\bea
\beta_{m_{11}^2} &=&\ldots\, +\,D\,\lambda_3 \left(\frac{}{}\left[3 \,\textrm{Tr}(\Delta_2 \Delta_2^\dagger) \,+\,
3 \,\textrm{Tr}(\Gamma_2 \Gamma_2^\dagger)\,+\,  \textrm{Tr}(\Pi_2 \Pi_2^\dagger)\right]\,m^2_{22}\right.
\nonumber \\
 & & \quad\quad\quad\quad\quad\left. -\,\left\{ \left[3 \,\textrm{Tr}(\Delta_2 \Delta_1^\dagger) \,+\, 3 \,\textrm{Tr}(\Gamma_2 \Gamma_1^\dagger)
\,+\, \textrm{Tr}(\Pi_2 \Pi_1^\dagger) \right]\,m^2_{12}\,+\, \textrm{h.c.} \right\}\frac{}{}\right)
\eea
and
\bea
\beta_{m_{22}^2} &=&\ldots\, +\,D\,\lambda_3 \left(\frac{}{}\left[3 \,\textrm{Tr}(\Delta_1 \Delta_1^\dagger) \,+\,
3 \,\textrm{Tr}(\Gamma_1 \Gamma1^\dagger)\,+\,  \textrm{Tr}(\Pi_1 \Pi_1^\dagger)\right]\,m^2_{11}\right.
\nonumber \\
 & & \quad\quad\quad\quad\quad\left. -\,\left\{ \left[3 \,\textrm{Tr}(\Delta_2 \Delta_1^\dagger) \,+\, 3 \,\textrm{Tr}(\Gamma_2 \Gamma_1^\dagger)
\,+\, \textrm{Tr}(\Pi_2 \Pi_1^\dagger) \right]\,m^2_{12}\,+\, \textrm{h.c.} \right\}\frac{}{}\right)\,,
\eea
and again we see that the terms in $m^2_{12}$ vanish and both contributions yield $\beta_{m_{11}^2 + m_{22}^2} \propto m_{11}^2 + m_{22}^2$.
The same will be valid for terms involving the couplings $\lambda_4$ and $\lambda_5$. The couplings $\lambda_6$ and $\lambda_7$ are a more amusing
situation, the former only contributing to $\beta_{m_{11}^2}$ and the latter only to $\beta_{m_{22}^2}$, in such a way that
\bea
\beta_{m_{11}^2} &=&\ldots\, +\,D\,\lambda_6 \left(\frac{}{}\left[3 \,\textrm{Tr}(\Delta_1 \Delta_2^\dagger) \,+\,
3 \,\textrm{Tr}(\Gamma_1 \Gamma_2^\dagger)\,+\,  \textrm{Tr}(\Pi_1 \Pi_2^\dagger) \,+\, \textrm{h.c.} \right]\,(m^2_{11}  +m^2_{22})\right.
\nonumber \\
 & & \quad\quad\quad\quad\quad\left. -\,\left\{ \left[3 \,\textrm{Tr}(\Delta_2 \Delta_2^\dagger) \,+\, 3 \,\textrm{Tr}(\Gamma_2 \Gamma_2^\dagger)
\,+\, \textrm{Tr}(\Pi_2 \Pi_2^\dagger) \right]\,m^2_{12}\,+\, \textrm{h.c.} \right\}\frac{}{}\right)
\eea
and
\bea
\beta_{m_{22}^2} &=&\ldots\, +\,D\,\lambda_7 \left(\frac{}{}\left[3 \,\textrm{Tr}(\Delta_2 \Delta_1^\dagger) \,+\,
3 \,\textrm{Tr}(\Gamma_2 \Gamma_1^\dagger)\,+\,  \textrm{Tr}(\Pi_2 \Pi_1^\dagger) \,+\, \textrm{h.c.} \right]\,(m^2_{11}  + m^2_{22})\right.
\nonumber \\
 & & \quad\quad\quad\quad\quad\left. -\,\left\{ \left[3 \,\textrm{Tr}(\Delta_1 \Delta_1^\dagger) \,+\, 3 \,\textrm{Tr}(\Gamma_1 \Gamma_1^\dagger)
\,+\, \textrm{Tr}(\Pi_1 \Pi_1^\dagger) \right]\,m^2_{12}\,+\, \textrm{h.c.} \right\}\frac{}{}\right)\,.
\eea
We see that now it is the terms proportional to $m^2_{11}$  and $m^2_{22}$ that vanish due to eq.~\eqref{eq:yiyj}, and the Yukawa coupling
structures multiplying $m^2_{12}$ are identical in both  equations above. This then leads to
\beq
\beta_{m_{11}^2 + m_{22}^2} \,=\, \ldots\, +\,D\,\left\{(\lambda_6 + \lambda_7)\,\left[ \textrm{Yukawa couplings}\right]\,m^2_{12}\,+\,
\textrm{h.c.} \right\}
\eeq
which of course is equal to zero since the $r_0$ symmetry implies $\lambda_7 = -\lambda_6$.

Therefore, all contributions to $\beta_{m_{11}^2 + m_{22}^2}$ involving Yukawa and quartic scalar couplings are proportional to
$m_{11}^2 + m_{22}^2$.

\begin{figure}[t]
\centering
\includegraphics[height=3cm,angle=0]{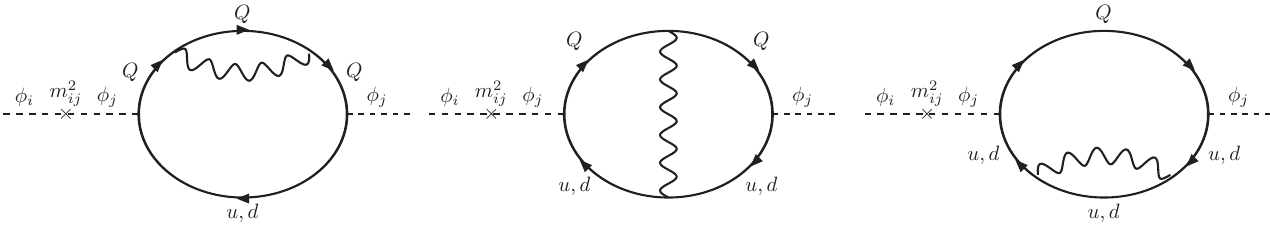}
\caption{Example of Feynman diagram contributing to the beta functions for the quadratic scalar coefficients involving
both Yukawa and gauge interactions. The ``$\times$" symbol
denotes a ``mass insertion" corresponding to the $m^2_{ij}$ coefficients.
}
\label{fig:diag3}
\end{figure}

\begin{itemize}
\item Yukawa and gauge coupling  contributions
\end{itemize}
Finally, the last contributions involve products of Yukawa and gauge couplings, and the demonstration is trivial: considering the mass insertions
possible in each case and the Yukawa structures allowed for each case (see Fig.~\ref{fig:diag3}), we will have
\bea
\beta_{m_{11}^2} &=&\ldots\, +\,G_1\, \left[3 \,\textrm{Tr}(\Delta_1 \Delta_1^\dagger) \,+\,
3 \,\textrm{Tr}(\Gamma_1 \Gamma_1^\dagger)\,+\,  \textrm{Tr}(\Pi_1 \Pi_1^\dagger)\right]\,m^2_{11}
\nonumber \\
 & &  \quad\quad -\,G_2 \left\{ \left[3 \,\textrm{Tr}(\Delta_1 \Delta_2^\dagger) \,+\, 3 \,\textrm{Tr}(\Gamma_1 \Gamma_2^\dagger)
\,+\, \textrm{Tr}(\Pi_1 \Pi_2^\dagger) \right]\,m^2_{12}\,+\, \textrm{h.c.} \right\}\,,
\eea
where we include all gauge, symmetry, pole factor in $G_1$ and $G_2$. For $\beta_{m_{22}^2}$ the result is quite simply
\bea
\beta_{m_{22}^2} &=&\ldots\, +\,G_1\, \left[3 \,\textrm{Tr}(\Delta_2 \Delta_2^\dagger) \,+\,
3 \,\textrm{Tr}(\Gamma_2 \Gamma_2^\dagger)\,+\,  \textrm{Tr}(\Pi_2 \Pi_2^\dagger)\right]\,m^2_{22}
\nonumber \\
 & & \quad\quad -\,G_2 \left\{ \left[3 \,\textrm{Tr}(\Delta_2 \Delta_1^\dagger) \,+\, 3 \,\textrm{Tr}(\Gamma_2 \Gamma_1^\dagger)
\,+\, \textrm{Tr}(\Pi_2 \Pi_1^\dagger) \right]\,m^2_{12}\,+\, \textrm{h.c.} \right\}\,.
\eea

Now, since $\Phi_1$ and $\Phi_2$ have exactly the same quantum numbers, gauge contributions to the beta functions of $m_{11}^2$ and
$m_{22}^2$ will per force be identical, which justifies the fact that the factors $G_1$ and $G_2$ are repeated in the above equations.
Eq.~\eqref{eq:yiyj} makes all terms proportional to $m^2_{12}$ vanish, and eq.~\eqref{eq:y2} makes the terms in square brackets
multiplying $G_1$ identical in both equations. Yet again, we obtain  $\beta_{m_{11}^2 + m_{22}^2} \propto m_{11}^2 + m_{22}^2$.

To conclude, when one considers the CP2 or CP3 Yukawa matrices of eqs.~\eqref{eq:yupcp2} and~\eqref{eq:yupcp3}, the beta functions
of the scalar squared mass coefficients are such that, at least to two-loop order, one has
\beq
\beta^{2L}_{m_{11}^2 + m_{22}^2} \,=\,\left[\textrm{Scalar, gauge, Yukawa couplings} \right]\, (m_{11}^2 + m_{22}^2)\,,
\eeq
so that the condition $m_{11}^2 + m_{22}^2 = 0$ is preserved under RG running\footnote{Provided the relations $\lambda_1 = \lambda_2$
and $\lambda_7 = -\lambda_6$, which complete the $r_0$ symmetry conditions, are also obeyed, of course.}.

\newpage
{\footnotesize
\bibliographystyle{utphys}
\bibliography{biblio}
}

\end{document}